\documentclass{nature}

\usepackage{acronym}
\usepackage{amsmath}
\usepackage{amssymb}
\usepackage{color}
\usepackage{hyperref}
\usepackage{graphicx}
\usepackage{times}
\usepackage{hyperref}
\usepackage{amssymb}
\usepackage{enumitem}
\usepackage{mathtools}
\usepackage{graphicx}
\usepackage{amssymb}
\usepackage{amsmath}
\usepackage{ulem}
\usepackage{epstopdf}
\usepackage{color}
\usepackage{graphicx}
\usepackage{xcolor}
\usepackage{textcomp}
\usepackage{units}
\usepackage{soul}
\usepackage{subfigure}
\usepackage{array}
\usepackage{comment}
\usepackage{multirow}
\newcolumntype{H}{>{\setbox0=\hbox\bgroup}c<{\egroup}@{}}
\newcommand{\aap}{Astron. and Astrophys.}
\newcommand{\apj}{Astrophys. J.}
\newcommand{\apjl}{Astrophys. J. Lett.}
\newcommand{\apjs}{Astrophys. J. Supp.}
\newcommand{\mnras}{Mon. Not. R. Astron. Soc.}
\newcommand{\nat}{Nat.}
\newcommand{\physrep}{Phys. Rep.}
\newcommand{\prd}{Phys. Rev. D}

\def\aap{Astron. Astrophys.}                
\newcommand{\aj}{Astron. J.}
\newcommand{\pasa}{Publ. Astron. Soc. Australia}

\newcommand{\prl}{Phys. Rev. Lett.}
\newcommand{\apss}{Astrophys. and Space Sci.}
\newcommand\sovast{{Soviet~Ast.}}

\def\araa{ARA\&A}  

\hypersetup{
    colorlinks=true,
    linkcolor=blue,
    filecolor=magenta,      
    urlcolor=cyan,
    citecolor=blue,
}
\usepackage{url}
\usepackage{listings}
\lstdefinestyle{mystyle}{
  commentstyle=\color{codegreen},
  keywordstyle=\color{magenta},
  numberstyle=\tiny\color{codegray},
  stringstyle=\color{codepurple},
  basicstyle=\footnotesize,
  breakatwhitespace=false,         
  breaklines=true,                 
  captionpos=b,                    
  keepspaces=true,                 
  numbers=left,                    
  numbersep=5pt,                  
  showspaces=false,                
  showstringspaces=false,
  showtabs=false,                  
  tabsize=2
}
\usepackage{graphicx}

\usepackage[affil-it]{authblk} % added 2020/03/19 

\begin{document}

\acrodef{BBH}{binary black hole}
\acrodef{BH}{black hole}
\acrodef{EM}{electromagnetic}
\acrodef{GW}{gravitational wave}
\acrodef{O1}{first observing}
\acrodef{PE}{parameter estimation}

\bibliographystyle{naturemag}

\title{The birth mass function of neutron stars}

\author{Zhi-Qiang You$^{1,2,3,4}$, Xingjiang Zhu$^{1,2,5}$\thanks{E-mail: zhuxj@bnu.edu.cn}, Xiaojin Liu$^{1,2,3}$, Bernhard Müller$^{6, 7}$, Alexander Heger$^{6, 7}$, Simon Stevenson$^{7, 8}$, Eric Thrane$^{6, 7}$, Zu-Cheng Chen$^{1,2,3,9,10}$, Ling Sun$^{7, 11}$, Paul Lasky$^{6, 7}$, Duncan K. Galloway$^{6, 7, 12}$, George Hobbs$^{13}$, Richard N. Manchester$^{13}$, He Gao$^{3, 5}$, Zong-Hong Zhu$^{3,14}$\thanks{E-mail: zhuzh@bnu.edu.cn}}

\maketitle

\begin{affiliations}
\item Department of Physics, Faculty of Arts and Sciences, Beijing Normal University, Zhuhai 519087, China
\item Advanced Institute of Natural Sciences, Beijing Normal University, Zhuhai 519087, China
\item School of Physics and Astronomy, Beijing Normal University, Beijing 100875, China
\item Henan Academy of Sciences, Zhengzhou 450046, Henan, China
\item Institute for Frontier in Astronomy and Astrophysics, Beijing Normal University, Beijing 102206, China
\item School of Physics and Astronomy, Monash University, Clayton, VIC 3800, Australia
\item OzGrav: The ARC Centre of Excellence for Gravitational Wave Discovery, Hawthorn, VIC 3122, Australia
\item Centre for Astrophysics and Supercomputing, Swinburne University of Technology, P.O. Box 218, Hawthorn, VIC 3122, Australia
\item Department of Physics and Synergetic Innovation Center for Quantum Effects and Applications, Hunan Normal University, Changsha, China.
\item Institute of Interdisciplinary Studies, Hunan Normal University, Changsha, China
\item Centre for Gravitational Astrophysics, College of Science, The Australian National University, Canberra, Australian Capital Territory, Australia.
\item Institute for Globally Distributed Open Research and Education (IGDORE), Gothenburg, Sweden.
\item Australia Telescope National Facility, CSIRO Space and Astronomy, Epping, New South Wales, Australia.
\item School of Physics and Technology, Wuhan University, Wuhan, Hubei 430072, China

\end{affiliations}
\clearpage

\begin{abstract}

The birth mass function of neutron stars encodes rich information about supernova explosions, double-star evolution and properties of matter under extreme conditions.  To date, it has remained poorly constrained by observations, however.
Applying probabilistic corrections to account for mass accreted by recycled pulsars in binary systems to mass measurements of 90 neutron stars, we find that the birth masses of neutron stars can be described by a unimodal distribution that smoothly turns on at $\mathbf{\unit[1.1]{\mathrm{M}_{\odot}}}$, peaks at $\mathbf{\sim \unit[1.27]{\mathrm{M}_{\odot}}}$, before declining as a steep power law.
Such a ``turn-on" power-law distribution is strongly favoured against the widely adopted empirical double-Gaussian model at the $\mathbf{3\sigma}$ level.
%, in agreement with expectations from the mass function of massive stars\cite{Kroupa01IMF,Schneider18IMF}.
The power-law shape may be inherited from the initial mass function of massive stars\cite{Kroupa01IMF,Schneider18IMF}, but the relative dearth of massive neutron stars implies that single stars with initial masses greater than $\mathbf{\sim \unit[18]{\mathrm{M}_{\odot}}}$ do not form neutron stars, in agreement with the absence of massive red supergiant progenitors of supernovae\cite{Smartt:2009MNRAS,Davies:2018MNRAS}.

\end{abstract}

The observational determination of mass function of neutron stars is a long-standing problem in astrophysics.
Early attempts were able to derive only loose constraints on the range of possible neutron star masses from a small number of measurements\cite{Joss76nat,Finn94prl}.
For a long time, all observed neutron-stars masses were in a narrow range, consistent with a Gaussian distribution with a mean of $\unit[1.35]{\mathrm{M}_{\odot}}$ and a width of $\unit[0.04]{\mathrm{M}_{\odot}}$ (ref.\cite{Thorsett99}).
Owning partly to the precise mass measurements of the Hulse-Taylor binary pulsar\cite{Taylor79Nat,Taylor92Nat}, a canonical mass of $\unit[1.4]{\mathrm{M}_{\odot}}$ has routinely been adopted in most current studies of neutron stars and in textbooks.

As the number of neutron-star mass measurements steadily increased, especially with the discoveries of massive neutron stars around $\unit[2]{\mathrm{M}_{\odot}}$ (refs.\cite{Demorest10Nat,Antoniadis13Sci}), it became apparent that the single-Gaussian model could no longer fit all available data.
Various groups have started to employ a two-Gaussian (2G) model to explain the observations, with the secondary mass peak shifting to higher values as more massive neutron stars have been measured\cite{Valentim11,Ozel12,Kiziltan13,Antoniadis16mass}.
The most widely adopted model consists of a narrow peak at $\unit[1.4]{\mathrm{M}_{\odot}}$ and a broad peak at $\unit[1.8]{\mathrm{M}_{\odot}}$ (ref.\cite{alsing2018}), and the two peaks are attributed to different formation channels and evolutionary histories of the neutron stars\cite{Horvath17ns}.

From a theoretical perspective, the neutron-star mass function is a powerful probe into supernova physics, compact binary evolution and the neutron-star equation of state.
It was once proposed that the two low-mass peaks (at $\unit[1.25]{\mathrm{M}_{\odot}}$ and $\unit[1.35]{\mathrm{M}_{\odot}}$) claimed in the mass distribution are the product of distinct supernova progenitor channels, electron-capture supernovae and low-mass iron-core-collapse supernovae\cite{Timmes96mass,Schwab10}.
An additional high-mass peak centred around $\unit[1.8]{\mathrm{M}_{\odot}}$ in the mass distribution was thought to arise due to the transition from convective central carbon burning to radiative central carbon burning\cite{Timmes96mass, sukhbold_14}.  Recent detailed studies, however, have revealed a complicated landscape of interacting shells with discontinuous changes in the core and remnant masses when the number of shells changes, or shells merge, as a function of initial or helium core mass\cite{MHLC_16,SWH_18}.

Efforts to link the observed mass distribution to neutron star formation channels are complicated because the fact that neutron stars in binary systems often go through a recycling process and gain some mass through accretion\cite{Alpar82Nat,RadhakSrini82,BhatVanden91,tlk12}.
Such a recycling process is the standard formation pathway for millisecond pulsars, which have by far dominated the population of neutron stars with precise mass measurements.
Although the recycling process may increase the neutron-star mass by up to $\sim0.3$~M$_\odot$ (refs.\cite{Alpar82Nat,tlk12, LI2021}), it has been argued that the high-mass peak around $\unit[1.8]{\mathrm{M}_{\odot}}$ in the observed mass spectrum is probably not caused by mass accretion but, instead, a hint of neutron stars being born massive\cite{Antoniadis16mass}.
Current supernova simulations, however, have difficulties in producing enough neutron stars around $\unit[1.8]{\mathrm{M}_{\odot}}$ as implied by the empirical 2G model\cite{Woosley20}.

%Understanding the neutron star mass function is becoming an increasingly important problem, largely owing to the rapid advancement of gravitational-wave astronomy.
%The Gravitational-Wave Transient Catalog 3 (GWTC-3) recently released by the LIGO-Virgo-KAGRA detector network contains 90 compact binary coalescences\cite{gwtc3}, including two binary neutron star mergers\cite{gw170817,GW190425}, two neutron star-black hole mergers\cite{2NSBH2021}, and a few more that might involve one neutron star.
%Several groups investigated the population properties of neutron stars observed in gravitational-wave mergers\cite{LandryRead21,Li21pmo,Zhujp21nsbh}; others outlined ideas to study the transition in the mass spectrum from neutron stars to black holes using gravitational-wave measurements\cite{Fishbach20matter,Farah21gap,Ye22gap}.
%Although the current gravitational-wave catalogue of neutron star binaries is too small to derive meaningful constraints on the mass function\cite{GWTC3pop}, linking mass measurements of neutron stars from gravitational waves with those of radio pulsars has already offered new insights into the population of binary neutron stars\cite{ZhuAshton20,shanika21dns}.

%\st{The number of neutron stars with mass estimates has grown significantly in the past few years.}
%\textcolor{red}{As an attempt to resolve the birth mass problem, we study the largest sample of neutron star mass measurements up to date. Specifically,} 

To determine the neutron-star mass distribution, we compiled a sample of 90 neutron stars for which well-determined mass estimates are available from observations of radio pulsars, gravitational waves and X-ray binaries.
For most of the neutron stars in our sample, the observed spin properties allowed us to securely categorize them into two subclasses: recycled and non-recycled (slowly rotating) neutron stars.
The first class is characterized by rapid spins (period $P\lesssim \unit[100]{ms}$), whereas the second class includes neutron stars with spin periods of the order of seconds or longer.
When constraining the birth-mass function of neutron stars, the key difference between the two subclasses is that observed masses of recycled pulsars need to be corrected for mass accreted throughout the recycling process, whereas the measured masses of slow neutron stars should equal their birth masses, as no mass-gaining process occurred.
See Section \ref{sec:data}, Extended Data Tables \ref{tab:psr_mass} and \ref{tab:psr_mass1} for details.

\begin{figure}
% \centering
\begin{center}
\includegraphics[width=115mm]{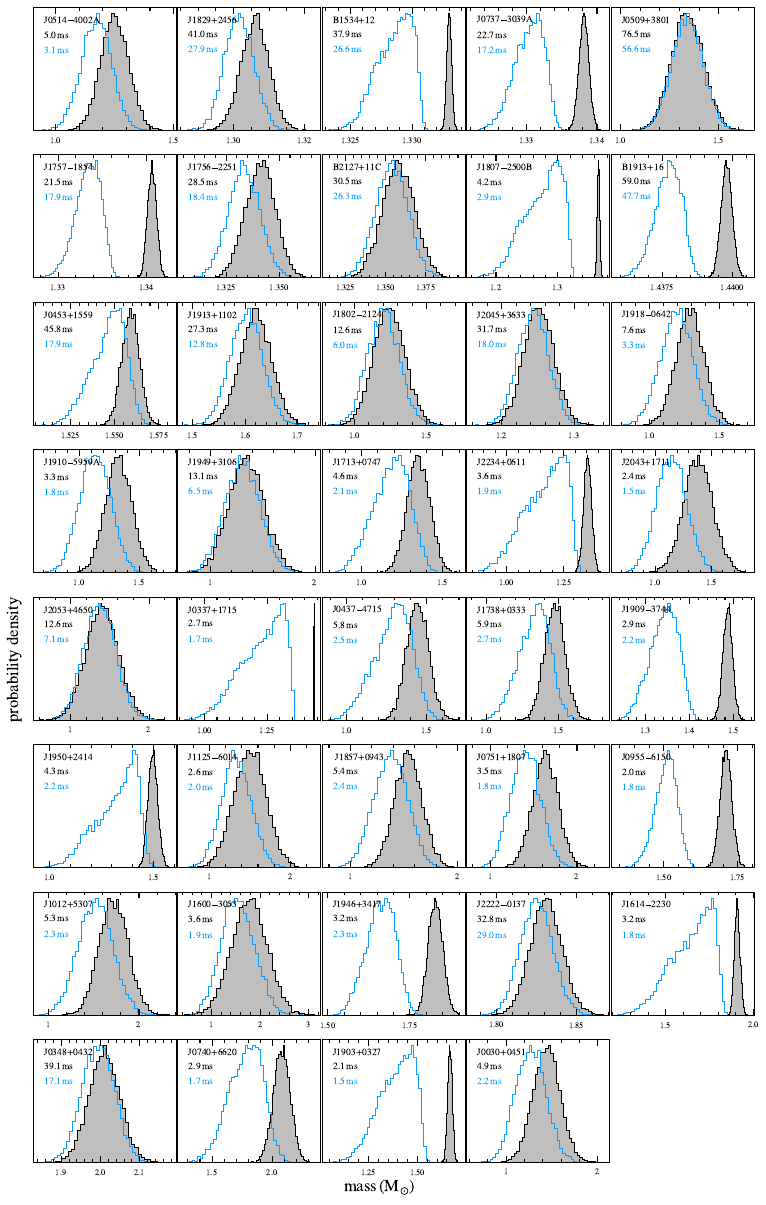}
% \includegraphics[height=.76\textheight]{plots/hist_39_mr_ana.jpg}
% \vspace{-5mm}
\\
\caption{\textbf{Individual mass distributions of 39 recycled pulsars}. Black lines with grey shaded area are measured masses, whereas blue lines are estimated birth masses after subtracting accreted masses based on the analytical prescription. Observed and plausible initial spin periods, along with pulsar names are given in each subplot. For readability, we scale different posteriors to the same ${y}$ coordinate height.}
\label{fig:mass_post}
\end{center}
\end{figure}

Figure \ref{fig:mass_post} shows the distributions of measured masses (black lines with grey-shaded areas) and estimated birth masses (blue lines) for 39 recycled pulsars.
In each subplot, we also list the pulsar name, observed spin period and plausible initial spin period.
The birth masses of recycled pulsars were estimated by subtracting accreted masses from their measured masses.
We adopted an analytical approach for estimating the accreted mass.
To the leading order, the total amount of accreted mass scales inversely with the final equilibrium spin period in the recycling process.
Based on the simple accretion spin-up theory\cite{Alpar82Nat,Lipunov84,tlk12}, we derived the probability distribution of the accreted mass by accounting for uncertainties in the accretion rate, neutron star moment of inertia, magnetic inclination angle and accretion disk-magnetosphere interactions.
See Section \ref{sec:model_deltam} and Extended Data Figs~\ref{fig:ppdot}, ~\ref{fig:DeltaM_spin} and ~\ref{fig:1614_mass}.

\begin{figure}
\begin{center}\includegraphics[width=103mm]{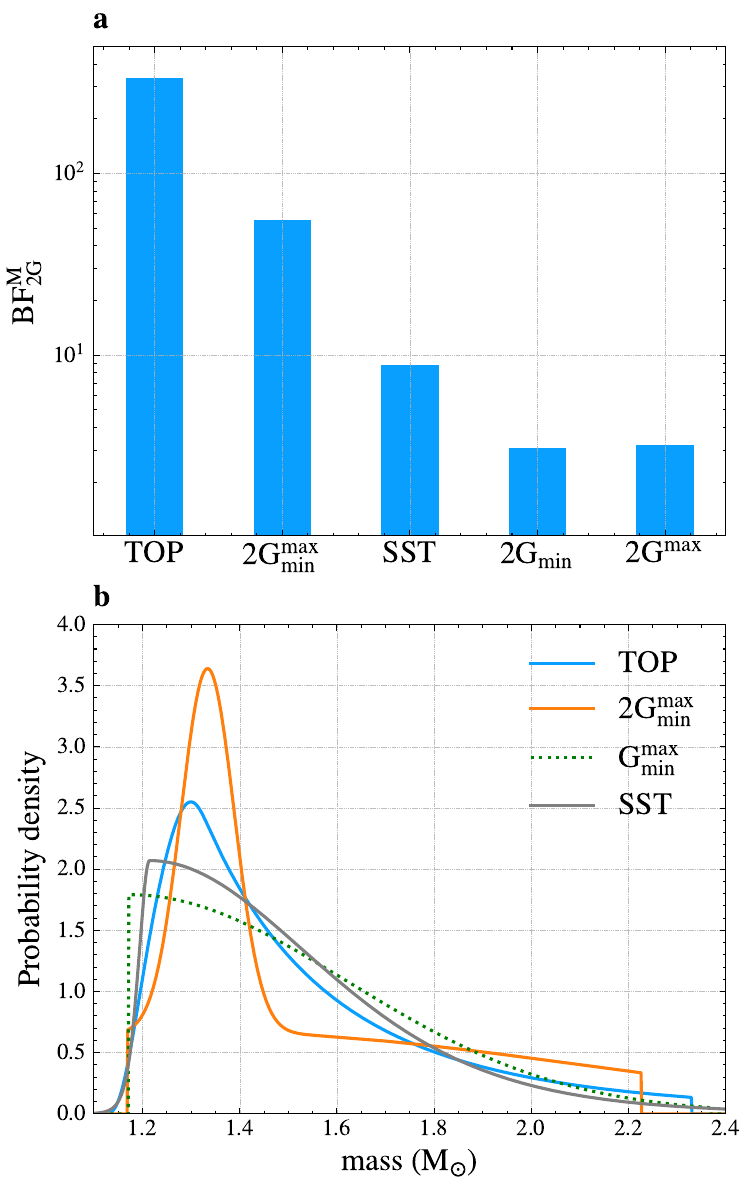}\\
\caption{
\textbf{Example mass models and their Bayes factors.} \textbf{a}, Bayes factor between various models (M) and the 2G model for the birth mass function of neutron stars. \textbf{b}, Mass models. In addition to the models listed in \textbf{a}, we also show a 
single-Gaussian distribution with both a low- and high-mass cutoff ($\rm G^{max}
_{min} $). Model parameters were set to their maximum-likelihood values inferred from data. SST, skewed Student’s $t$-distribution.
}
\label{fig:BF_2panels}
\end{center}
\end{figure}

We fitted a variety of models to the mass distribution of neutron stars.
The Gaussian models are a 2G distribution with both a low- and high-mass cutoff ($\rm 2G^{max}_{min}$ ), a 2G distribution with a low-mass cutoff ($\rm 2G_{min}$) and a 2G distribution with a high-mass cutoff ($\rm 2G^{max}$).
In addition to the Gaussian-family models, 
we considered a turn-on power-law(TOP) distribution, which allows an exponential turn-on from the minimum mass\cite{ColmEric18bh}, as well as a skewed Student's $t$-distribution.
Some of the models are illustrated in panel \textbf{b} of Figure \ref{fig:BF_2panels}.
Bayesian model selection was performed to find the model that best describes the data.
The Bayes factors, which quantify the statistical support for one model against another, between five top-performing models and the empirical two-Gaussian (2G) model are shown in panel \textbf{a} of Figure \ref{fig:BF_2panels}.
The best-performing model was the turn-on power law (TOP), which is preferred against the 2G model with a Bayes factor greater than 300.
Assuming equal prior probabilities for both models, this result means an odds of $0.003$, or equivalently $\sim 3\sigma$ significance. Potential selection effects are discussed in Section \ref{sec:disc} of Methods, and further details can be found in Extended Data Figure~\ref{fig:mass-spin}.

\begin{figure}
\begin{center}
  \includegraphics[width=130mm]{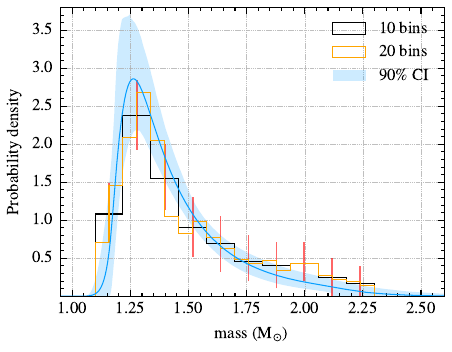} \\ 
% \vspace{-3mm}  
  \caption{
\textbf{Reconstructed birth-mass function of neutron stars based on the preferred TOP model.} 
The solid blue line shows the population predictive distribution, whereas the shaded blue band indicates the 90\% credible interval. 
Also plotted are a ten-bin histogram (mean value) with 90\% credibility error bars and a 20-bin histogram (mean value) for the data, showing the robustness of our model in fitting the underlying distribution. CI, confidence interval.}
\label{fig:ppd_hist}
\end{center}
\end{figure}

\begin{figure}
\begin{center}
  \includegraphics[width=140mm]{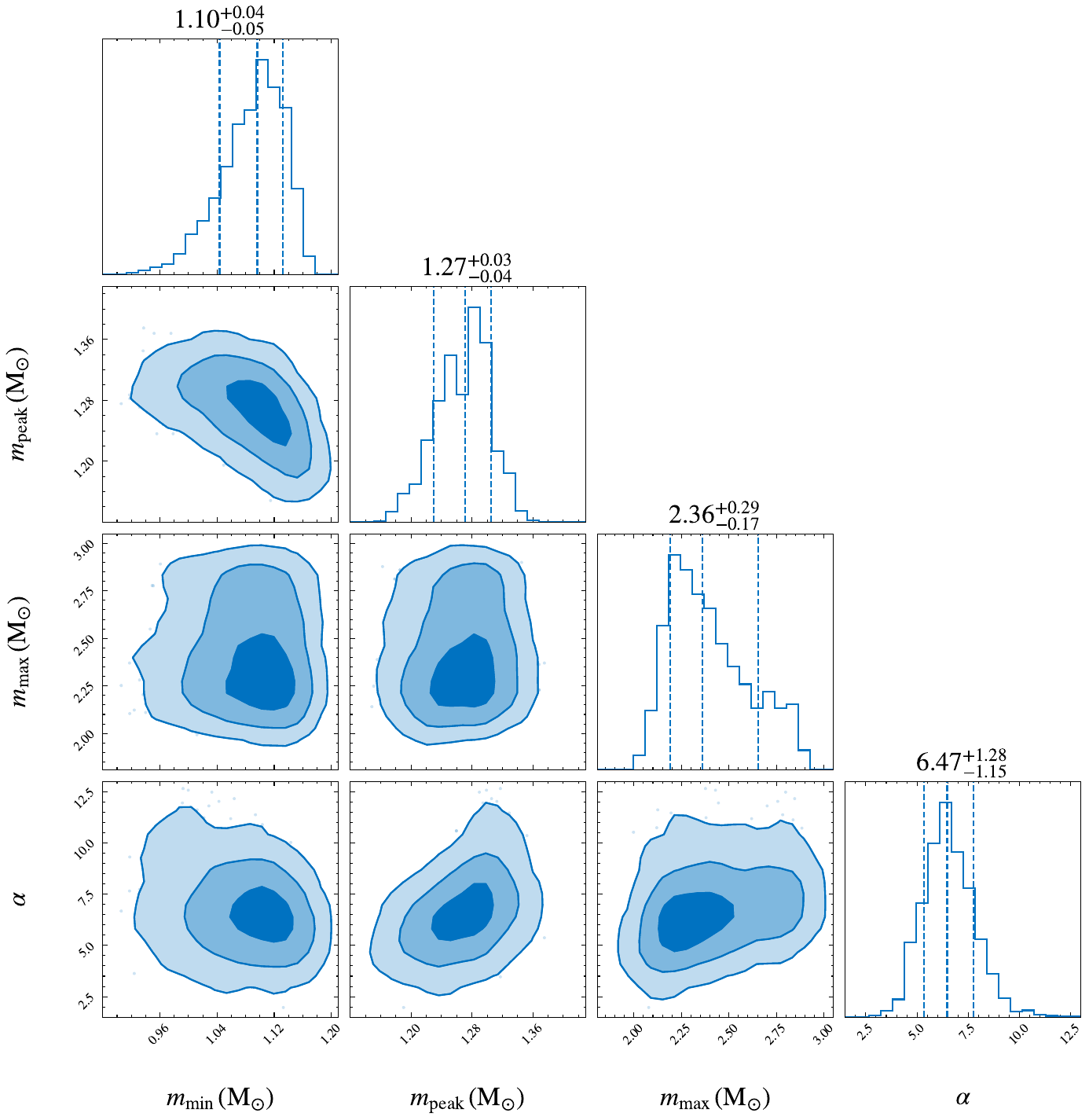} \\
%\vspace{-3mm}  
  \caption{\textbf{Posterior distribution of the TOP model parameters}. Shown are the minimum neutron-star mass $m_{\rm min}$, the peak of the mass function $m_{\rm peak}$, the maximum mass $m_{\rm max}$ and the power-law index $\alpha$.
}
\label{fig:TOP_m_peak}
\end{center}
\end{figure}

Figure \ref{fig:ppd_hist} shows the reconstructed birth-mass function of neutron star for the turn-on power law model, as well as histograms for two different bin sizes.
We found that the minimum mass was $\unit[1.1^{+0.04}_{-0.05}]{\mathrm{M}_{\odot}}$. The distribution peaked at $\unit[1.27^{+0.03}_{-0.04}]{\mathrm{M}_{\odot}}$ and declined with a power-law index of $6.5^{+1.3}_{-1.2}$.
We found marginal evidence for a maximum-mass cutoff, with a Bayes factor of 2.
The maximum mass was loosely constrained to be $\unit[2.36^{+0.29}_{-0.17}]{\mathrm{M}_{\odot}}$.
(All numbers quoted are median and $1\sigma$ credible intervals.)
The posterior distributions of model parameters are presented in Figure~\ref{fig:TOP_m_peak}.
We found no evidence for other peaks on top of the power-law function, although such a possibility cannot be ruled out either.

The final evolution of the lowest-mass stars that make iron-core-collapse supernovae is notably diverse and a challenge to model due to the complicated physics\cite{WoosleyHeger_15}.  
Hence, any constraints from observations on this mass regime are invaluable.  
Such low-mass iron-core-collapse supernovae are predicted to produce neutron stars with masses as low as or potentially below $1.15\,\mathrm{M}_\odot$ (refs.\cite{Timmes96mass, WoosleyHeger_15}), consistent with our observed minimum.  
Electron-capture supernovae are unlikely to produce such low-mass neutron stars\cite{huedepohl_10}.  They, however, may make neutron stars with masses within the peak of the distribution.
The data presently do not provide support for the two discrete low-mass peaks that are expected from theoretical calculations\cite{Schwab10}. Instead, the distribution is probably a superposition of several overlapping discontinuous (as a function of initial mass) contributions\cite{MHLC_16,SWH_18}, overlaid with some scatter due to variations in the turbulent supernova explosions\cite{KEJ_21}.  
Particularly marked is the absence of the second peak predicted  at $\unit[1.35]{\mathrm{M}_{\odot}}$, which probably hints at the larger importance of lower-mass core-collapse and of stripped-core supernova progenitors\cite{2020ApJ...890...51E}.

%Additionally, if the small peak at $\sim\unit[2]{\mathrm{M}_{\odot}}$ (Figure~\ref{fig:ppd_hist}) is a true feature of the neutron star mass distribution, it may be related to the predicted \emph{island of explodability} for initial stellar masses above $\sim\unit[23]{\mathrm{M}_{\odot}}$\cite{MHLC_16}.
% \bm{The next few sentence is contradicts our new interpretation of the data and should eventually be removed. This also need to be stated in the reply letter}
% The smooth tail of the distribution up to $\unit[2.4]{\mathrm{M}_{\odot}}$ indicates successful supernova explosions up to higher iron-silicon core mass and progenitor compactness than predicted by current parameterised supernova models \cite{sukhbold_16,MHLC_16,2020ApJ...890...51E}. 
If we accept the core structures predicted by current stellar evolution models, the absence of prominent peaks and the measured power-law slope of the neutron-star mass distribution have important implications for supernova explosion physics. Specifically, the most parsimonious explanation of TOP is that neutron stars %made by more massive neutron stars
from more massive progenitors undergo substantial net accretion during the explosion and that explosions occur only for stars with main sequence masses of 
$\lesssim \unit[18]{\mathrm{M}_{\odot}}$ in agreement with the observed absence of massive red supergiants as progenitors for core-collapse supernovae\cite{Smartt:2009MNRAS,Davies:2018MNRAS}
or for stripped stars with similar core structures.
We demonstrate this through the use of semi-analytic supernova models\cite{MHLC_16}, as described in Methods Section~\ref{sec:supernova_modeling}.

Knowing the neutron-star mass function is also essential for understanding the population of binary neutron star mergers observed by their gravitational wave signal. A fraction of these mergers result in stable very high-mass neutron stars.
It also allows us to obtain a more accurate prediction of the frequency band for post-merger gravitational-wave signals, which can substantially benefit the design 
and optimization of future-generation gravitational-wave observatories that aim to probe post-merger physics\cite{CEHS2021,Srivastava2022,Martynov2019}.
The remnant compact object formed in a binary neutron star merger is in an extreme, high-density, high-energy state and encodes unique information about nuclear matter that the progenitor stars and the inspiral phase do not provide. 
Observations of the post-merger gravitational waves may provide unprecedented opportunities for exploring hot equations of state at supranuclear densities and identifying phase transitions in extreme nuclear matter\cite{Takami2015,Rezzolla2016,Bauswein2019}. 
The post-merger emissions are at frequencies mainly determined by the progenitor neutron-star masses and the yet unknown nuclear equation of state, spanning a wide range roughly from $\unit[1]{}$ to 
$\unit[4]{kHz}$ (refs.\cite{Shibata2006,Baiotti2008}).

%What does it mean for supernova physics?
%Implications for gravitational-wave astronomy and neutron star studies.

\renewcommand{\baselinestretch}{1.0}

\clearpage

\clearpage

\clearpage
\begin{addendum}
\item[Data Availability] 
%All the neutron star mass measurements used in this study are presented in Extended Data Tables 1 and 2, with original references properly cited. These mass measurements, accreted-mass corrections for recycled pulsars, posterior samples from Bayesian inference, data behind the tables and figures can be accessed at \url{https://doi.org/10.5281/zenodo.14375273}.
All the neutron-star mass measurements used in this study are listed in Extended Data Tables 1 and 2 with the original references. These mass measurements, accreted-mass corrections for recycled pulsars, posterior samples from Bayesian inference and the data behind Extended Data Tables 1 and 2 and Figures 1–4 and Extended Data Figures 1–4 are available via Zenodo at \url{https://doi.org/10.5281/zenodo.14375273}. Source data are provided with this paper.

\item[Code Availability]
The following open-source software packages were used in this paper: \href{https://github.com/pathakdhruv/GalDynPsr}{GalDynPsr}, \href{https://github.com/ColmTalbot/gwpopulation}{gwpopulation}, \href{https://github.com/bilby-dev/bilby}{BILBY} and \href{https://github.com/joshspeagle/dynesty}{dynesty}.
The Python scripts used for data analysis and figure generation are publicly available via GitHub at \url{https://github.com/GW-BNUZ/NSbirthMass}.
 
%{\item We thank Zhenwei Li, Xuefei Chen and Zhanwen Han for providing data from their numerical simulations that are plotted in Extended Data Figure~\ref{fig:DeltaM_spin}. We also thank Ilya Mandel and Matthew Bailes for useful discussions.
%This work was supported by the National Natural Science Foundation of China under grant numbers~12203004 (X.Z.), 12021003 and 12433001 (Z.-H.Z.), 12305059 (Z.-Q.Y.), 12405056 (Z.-C.C.).
%X.Z. and H.G. are supported by the Fundamental Research Funds for the Central Universities.
%This work was supported in part by the Australian Research Council (ARC) Centre of Excellence for Gravitational Wave Discovery (OzGrav), through project numbers CE170100004 and CE230100016..
%S.S. is a recipient of an ARC Discovery Early Career Research Award (DE220100241). This work was supported by ARC Discovery grants DP240101786 (B.M., A.H.) and DP240103174 (A.H.).}

\item We thank Z. Li, X. Chen and Z. Han for providing data from their numerical simulations, which are plotted in Extended Data Figure~\ref{fig:DeltaM_spin}. We also thank I. Mandel and M. Bailes for useful discussions. This work was supported by the National Natural Science Foundation of China (Grant Nos. 12203004 to X.Z., 12021003 and 12433001 to Z.-H.Z., 12305059 to Z.-Q.Y. and 12405056 to Z.-C.C.). X.Z. and H.G. are supported by the Fundamental Research Funds for the Central Universities. This work was supported in part by the Australian Research Council (ARC) Centre of Excellence for Gravitational Wave Discovery (Project Nos. CE170100004 and CE230100016). S.S. is a recipient of an ARC Discovery Early Career Research Award (DE220100241). This work was supported by ARC Discovery Grants DP240101786 (to B.M. and A.H.) and DP240103174 (to A.H.).

\item[Author Contributions] All authors contributed to the work
  presented in this paper.
\item[Competing Interests] The authors declare no competing interests.
% \item[Author Information] Reprints and permissions information is
%   available at \url{http://www.nature.com/reprints}.  The authors
%   declare no competing financial interests.  Correspondence and
%   requests for materials should be addressed to
%   \href{mailto:zhuxj@bnu.edu.cn}{zhuxj@bnu.edu.cn} and \href{mailto:zhuzh@bnu.edu.cn}{zhuzh@bnu.edu.cn}.
\end{addendum}

\clearpage

\begin{methods}
\renewcommand{\baselinestretch}{1.0}
\selectfont

% Reset the Figure Counter
\setcounter{figure}{0}
\renewcommand{\figurename}{Extended Data Figure}
\renewcommand{\tablename}{Extended Data Table}

\section{Mass measurements of neutron stars}
\label{sec:data}

Extended Data Tables \ref{tab:psr_mass} and \ref{tab:psr_mass1} list measurements of the mass of 90 neutron stars compiled from the literature.
These include 13 double neutron star systems, 27 neutron star and white dwarf binaries, three neutron star and main sequence star binaries, 12 ``spider" binary pulsar systems (six redback and six black widow binaries), one isolated neutron star, 14 X-ray binaries (11 high-mass and three low-mass X-ray binaries), five gravitational-wave events of compact binary mergers (two binary neutron star and three neutron star and black hole systems).

A key aspect of this work is that we divided neutron stars into two subclasses: recycled and non-recycled (slow) neutron stars.
For most neutron stars, the classification is based on measured spin properties; recycled neutron stars have period $P \lesssim \unit[100]{ms}$ and $\dot{P} \lesssim 10^{-17}$, whereas slow neutron stars have spin periods of the order of seconds or longer. For the latter, no notable accretion-induced spin-up has yet occurred.
Moreover, we made the following general assumptions.
First, for double neutron star systems and binary neutron star mergers, we assumed that each consists of a recycled and a slow neutron star, with masses $m_r$ and $m_s$, based on the standard formation theory as represented by the double pulsar system PSR J0737$-$3039A/B\cite{Tauris17bns}.
Second, neutron stars in high-mass X-ray binaries are considered to be slow neutron stars; these are neutron stars in the early stage of the recycling process and our assumption is effectively that the amount of accreted mass is negligible.
Third, we assumed that neutron stars in neutron star and black hole mergers are non-recycled. This is believed to be the prevailing scenario in which the neutron star is the second object formed in the standard isolated binary formation channel\cite{Debatri21_nsbh,Floor21a}.

%We restrict ourselves to radio pulsar timing observations to make a homogenous data set with minimized systematics. We choose to do so also because pulsar observations provide the highest measurement precision and account for the majority of currently available data, and more importantly, the observed characteristics of recycled pulsars enable the corrections of accreted mass and thus allow us to probe the birth mass function of neutron stars. We compile a list of mass measurements with a $1\sigma$ uncertainty below $\unit[0.2]{\mathrm{M}_{\odot}}$ from the literature. These are all precise measurements for which the mass posteriors can be approximated by Gaussian functions.

Our division results in 56 recycled neutron stars and 34 slow neutron stars. For the two recycled neutron stars in GW170817 and GW190425, and the unseen companion of PSR J1906+0746 (assumed to be a recycled neutron star), neither $P$ nor $\dot{P}$ have been measured. Therefore, we did not include these three $m_r$ measurements when accounting for the accreted mass in inferring their birth masses.
PSRs J0514$-$4002A, B2127+11C and J1807$-$2500B are in globular clusters, which means that they may have originated from a different dynamical formation channel\cite{PhinneySigurd91,Grindlay06,Lee10sgrb,Ye20GCbns}.
Nevertheless, we included their companions in the category of slow neutron stars (which is a possibility even for the non-standard dynamical formation channel).
Third, it was argued that the companion of PSR J0453+1559 could be a white dwarf born in a thermonuclear electron-capture supernova\cite{TaurisJ0453}.
We do not expect our analysis results to be substantially affected by the inclusion of these four $m_s$ measurements (companions for three globular-cluster pulsars and PSR J0453+1559).

The input for our hierarchical Bayesian analysis (described in Section \ref{sec:bayes} in Methods) are individual mass posteriors. For mass measurements reported with symmetric errors in original publications, they are approximated by Gaussian functions; for gravitational-wave events, we used posterior samples that are publicly available in the relevant publications\cite{gw170817,GW190425,Zhu_GW170817,Abbott_GW200115,Abbott_GW191219_GWTC3}.

%\clearpage

\begin{table}
\centering
\caption{\label{tab:psr_mass}\textbf{Mass measurements of neutron stars in double neutron star and neutron star-white dwarf systems.} Listed here are pulsar name, the mass of recycled neutron star ($m_{\rm r}$), slow neutron star ($m_{\rm s}$) and companion star ($m_{\rm c}$), spin period ($P$), the intrinsic spin-down rate ($\dot{P}$), orbital period ($P_{\rm b}$) and eccentricity ($e_0$). Mass uncertainties are quoted in parentheses.
%Note that PSR J1903+0327 has a main-sequence star companion which apparently could not have recycled the millisecond pulsar; the donor star responsible for pulsar recycling was lost somewhere\cite{Freire11J1903origin}. 
%The first 12 are double neutron star systems and we assume each consists of a recycled and a normal (slow) neutron star following the standard formation scenario as represented by the double pulsar system (PSR J0737$-$3039A/B). For $m_r$ and $m_s$, figures in parentheses are $1\sigma$ uncertainties in the last quoted digit.
}
\resizebox{\textwidth}{!}{
\begin{tabular}
{lcccccccc}
 \hline
  \hline
  Pulsar Name & $m_{\rm r}$ ($\mathrm{M}_{\odot}$) & $m_{\rm s}$ ($\mathrm{M}_{\odot}$) & $m_{\rm c}$ ($\mathrm{M}_{\odot}$) & $P$ (ms) & $\dot{P}\,(10^{-18})$ & $P_{\rm b}$ (day) & $e_0$ & Reference\\
  \hline
  \multicolumn{9}{c}{Double neutron star systems} \\
  J0514$-$4002A & 1.25(6) & 1.22(6) & -- & 4.99 & 0.084 & 18.785 & 0.888 & \cite{Ridolfi19_J0514a} \\
  J1829$+$2456  & 1.306(4) & 1.299(4) & -- & 41.01 & 0.0435 & 1.176 & 0.139 & \cite{PSRJ1829mass}\\
  J1906$+$0746  & 1.322(11) & 1.291(11) & -- & 144.1 & 20268 & 0.166 & 0.085 & \cite{PSR1906}\\
  B1534$+$12 & 1.3330(2) & 1.3455(2) & -- & 37.90 & 2.364 & 0.421 & 0.274 & \cite{FonsecaB1534}\\
  J0737$-$3039A  & 1.3381(7) & 1.2489(7) & -- & 22.70 & 1.761 & 0.102 & 0.088 & \cite{Kramer06Sci}\\
  J0509$+$3801 & 1.34(8) & 1.46(8) & -- & 76.54 & 7.924 & 0.380 & 0.586 & \cite{Lynch0509}\\
  J1757$-$1854 & 1.3406(5) & 1.3922(5) & -- & 21.50 & 2.654 & 0.184 & 0.606 & \cite{Cameron22_PSRJ1757} \\
  J1756$-$2251  & 1.341(7) & 1.230(7) & -- & 28.46 & 1.015 & 0.320 & 0.181 & \cite{FerdmanPSR1756}\\
  B2127$+$11C  & 1.358(10) & 1.354(10) & -- & 30.53 & 5.01 & 0.335 & 0.681 & \cite{Jacoby06}\\
  J1807$-$2500B  & 1.3655(21) & 1.2064(21) & -- & 4.19 & 0.113 & 9.957 & 0.747 & \cite{Lynch12}\\
  B1913$+$16  & 1.4398(2) & 1.3886(2) & -- & 59.03 & 8.645 & 0.323 & 0.617 & \cite{Weisberg10}\\
  J0453$+$1559  & 1.559(5) & 1.174(4) & -- & 45.78 & 0.177 & 4.072  & 0.113 & \cite{Martinez15}\\
  J1913$+$1102 & 1.62(3) & 1.27(3) & -- & 27.29 & 0.176 & 0.206 & 0.090 & \cite{Ferdman20} \\
  \hline
  \multicolumn{9}{c}{Neutron star-white dwarf systems} \\
  J1802$-$2124 & 1.24(11) & -- & 0.78 & 12.65 & 0.071 & 0.699 & $10^{-6}$ & \cite{Ferdman10J1802mass} \\
  J2045$+$3633 & 1.251(21) & -- & 0.873 & 31.68 & 0.594 & 32.30 & 0.017 & \cite{J2045mass} \\
  J1141$-$6545 & -- & 1.27(1) & 1.01 & 393.9 & 4295 & 0.198 & 0.172 & \cite{Vivek20Sci1141} \\
  J1918$-$0642 & 1.29(10) & -- & 0.231 & 7.646 & 0.024 & 10.91 & $10^{-5}$ & \cite{NANOGrav11yr} \\
  J1910$-$5959A & 1.33(11) & -- & 0.18 & 3.266 & 0.0056 & 0.837 & $10^{-6}$ & \cite{Corongiu12_J1910} \\
  J1949$+$3106 & 1.34(17) & -- & 0.81 & 13.14 & 0.094 & 1.95 & $10^{-5}$ & \cite{ZhuWW19} \\
  J1713$+$0747 & 1.35(7) & -- & 0.292 & 4.57 & 0.008 & 67.83 & $10^{-4}$ & \cite{NANOGrav11yr} \\
  J2234$+$0611 & 1.353(17) & -- & 0.298 & 3.577 & 0.006 & 32.0 & 0.129 & \cite{Stovall19_J2234} \\
  J2305$+$4707 & -- & 1.38(10) & 1.26 & 1066 & 569 & 12.34 & 0.658 & \cite{Thorsett99,Kerkwijk99J2305} \\
  J2043$+$1711 & 1.38(13) & -- & 0.173 & 2.38 & 0.004 & 1.48 & $10^{-6}$ & \cite{NANOGrav11yr} \\
  J2053$+$4650 & 1.40(21) & -- & 0.86 & 12.59 & 0.166 & 2.45 & $10^{-5}$ & \cite{PSR_J2053mass} \\
  J0337$+$1715 & 1.4359(3) & -- & 0.1973 & 2.733 & 0.017 & 1.629 & $10^{-3}$ & \cite{Archibald18_PSRtriple} \\
  J0437$-$4715 & 1.44(7) & -- & 0.224 & 5.76 & 0.014 & 5.74 & $10^{-5}$ & \cite{PPTAdr1e_timing} \\
  J1738$+$0333 & 1.47(7) & -- & 0.181 & 5.85 & 0.022 & 0.355 & $10^{-7}$ & \cite{Anton12_J1738mass} \\
  J1909$-$3744 & 1.486(11) & -- & 0.208 & 2.95 & 0.0027 & 1.533 & $10^{-7}$ & \cite{PPTAdr2timing} \\
  J1950$+$2414 & 1.496(23) & -- & 0.28 & 4.30 & 0.020 & 22.19 & 0.08 & \cite{ZhuWW19} \\
  J1125$-$6014 & 1.5(2) & -- & 0.31 & 2.63 & 0.002 & 8.753 & $10^{-6}$ & \cite{PPTAdr2timing} \\
  J1857$+$0943 & 1.54(13) & -- & 0.263 & 5.36 & 0.017 & 12.327 & $10^{-5}$ & \cite{PPTAdr2timing} \\
  J0751$+$1807 & 1.64(15) & -- & 0.16 & 3.48 & 0.006 & 0.263 & $10^{-6}$ & \cite{EPTA16_Desvignes_42p} \\
  J0955$-$6150 & 1.71(3) & -- & 0.254 & 2.0 & 0.15 & 24.58 & 0.118 & \cite{Serylak22J0955} \\
  J1012$+$5307 & 1.72(16) & -- & 0.165 & 5.26 & 0.011 & 0.605 & $10^{-6}$ & \cite{Sanchez20_J1012mass} \\
  J1600$-$3053 & 1.77(36) & -- & 0.34 & 3.598 & 0.009 & 14.35 & $10^{-3}$ & \cite{PPTAdr2timing} \\
  J1946$+$3417 & 1.828(22) & -- & 0.2556 & 3.17 & 0.003 & 27.02 & 0.134 & \cite{Barr17_J1946} \\
  J2222$-$0137 & 1.831(10) & -- & 1.319 & 32.82 & 0.017 & 2.446 & $10^{-3}$ & \cite{GuoYJ21} \\
  J1614$-$2230 & 1.908(16) & -- & 0.493 & 3.15 & 0.0096 & 8.687 & $10^{-6}$ & \cite{NANOGrav11yr} \\
  J0348$+$0432 & 2.01(4) & -- & 0.172 & 39.12 & 0.241 & 0.102 & $10^{-6}$ & \cite{Antoniadis13Sci} \\
  J0740$+$6620 & 2.08(7) & -- & 0.253 & 2.89 & 0.0054 & 4.77 & $10^{-6}$ & \cite{FonsecaJ0740mass} \\
%  B1516+02B  & 2.08(19) & -- & 0.15 & 7.95 & $-0.003$ & 6.858 & 0.138 & \cite{Freire08M5} \\
  \hline
\hline 
\end{tabular}
}
\end{table}

%\clearpage
\renewcommand{\arraystretch}{0.94}
\begin{table}
\centering
\caption{\label{tab:psr_mass1} 
\textbf{Mass measurements of neutron stars in other types of systems.} Entries filled with a dash line are inapplicable, while leaving blank indicates an unknown parameter. Error ranges for gravitational wave events represent 1$\sigma$ errors.}
%The same as Extended Data Table~\ref{tab:psr_mass}, but for neutron stars in other types of systems. Entries filled with a dash line are inapplicable, while leaving blank indicates an unknown parameter. Error ranges for Gravitational wave events represent 1-$\sigma$ errors.}
%\vspace{-0.1cm}
\resizebox{\textwidth}{!}{
\begin{tabular}
{lccccccc}
 \hline
  \hline
  Name & $m_{\rm r}$ ($\mathrm{M}_{\odot}$) & $m_{\rm s}$ ($\mathrm{M}_{\odot}$) & $m_{\rm c}$ ($\mathrm{M}_{\odot}$) & $P$ (ms) & $\dot{P}\,(10^{-18})$ & $P_{\rm b}$ (day) & Reference\\
  \hline
  \multicolumn{8}{c}{Neutron star plus main-sequence star systems} \\
  LAMOST J1123 & -- & 1.24(3) & 0.61 &  &  & 0.274 & \cite{LAMOST1123}\\
  PSR J0045$-$7319 & -- & 1.58(34) & 10 & 926.3 & 4463 & 51.17 & \cite{NANOGrav11yr}\\
  PSR J1903$+$0327 & 1.666(12) & -- & 1.033 & 2.15 & 0.0188 & 95.17 & \cite{Thorsett99} \\
  
  \hline
  \multicolumn{8}{c}{Redback binary pulsar systems} \\
  PSR J1723$-$2837 & 1.22(26) & -- & 0.36 & 1.856 & 0.008 & 0.615 & \cite{Strader_J1723} \\
  PSR J2039$-$5617 & 1.3(1) & -- & 0.2 & 2.651 & 0.014 & 0.228 & \cite{Clark_J2039} \\
  PSR J2339$-$0533 & 1.64(27) & -- & 0.35 & 2.884 & 0.014 & 0.193 & \cite{Strader_J1723} \\
  PSR J1023$+$0038 & 1.71(16) & -- & 0.24 & 1.688 & 0.0053 & 0.198 & \cite{Deller12_J1023} \\
  PSR J2129$-$0429 & 1.74(18) & -- & 0.44 & 7.62 &  & 0.635 & \cite{Bellm_J2129} \\
  PSR J2215$+$5135 & 2.28(10) & -- & 0.25 & 2.610 & 0.0334 & 0.173 & \cite{Kandel_J2215} \\
  \hline
  \multicolumn{8}{c}{Black widow binary pulsar systems} \\
  PSR J1555$-$2908 & 1.67(7) & -- & 0.06 & 1.788 & 0.045 & 0.234 & \cite{Kennedy_J1555} \\
  PSR J1301$+$0833 & 1.74(20) & -- & 0.035 & 1.84 &  & 0.27 & \cite{Romani_J1301} \\
  PSR J1810$+$1744 & 2.13(4) & -- & 0.065 & 1.66 &  & 0.15 & \cite{Romani_J1810} \\
  PSR J1653$-$0158 & 2.17(21) & -- & 0.014 & 1.968 & 0.0008 & 0.052 & \cite{Nieder_J1653} \\
  PSR J1959$+$2048 & 2.18(9) & -- & 0.033 & 1.607 & 0.0169 & 0.382 & \cite{Kandel_J2215} \\
  PSR J0952$-$0607 & 2.35(17) & -- & 0.032 & 1.41 & 0.0046 & 0.268 & \cite{Romani_J0952} \\
\hline
  \multicolumn{8}{c}{Isolated neutron stars} \\
  PSR J0030$+$0451 & 1.44(15) & -- & -- & 4.865 & 0.010 & -- & \cite{Miller_J0030,RileyJ0030} \\
  \hline
  \multicolumn{8}{c}{Low-mass X-ray binaries} \\
  4U 1608$-$52 & 1.57(30) & -- &  & 1.61 &  & 0.537 & \cite{Ozel_4U1608} \\
  KS 1731$-$260 & 1.61(37) & -- &  & 1.91 &  &  & \cite{Ozel_4U1608} \\
  X1822$-$371 & -- & 1.69(13) & 0.46 & 590 &  & 0.232 & \cite{Iaria_2015AA} \\
\hline
  \multicolumn{8}{c}{High-mass X-ray binaries (spin periods in seconds)} \\
  4U 1538$-$522 & -- & 1.02(17) & 16 & 526.8 &  & 3.728 & \cite{Iaria_2015AA} \\
  SMC X$-$1 & -- & 1.21(12) & 18 & 0.71 &  & 3.892 & \cite{10_hmxb} \\
  XTE J1855-026 & -- & 1.41(24) & 21 & 360.7 &  & 6.074 & \cite{10_hmxb} \\
  LMC X$-$4 & -- & 1.57(11) & 18 & 13.5 &  & 1.408 & \cite{10_hmxb} \\
  Cen X$-$3 & -- & 1.57(16) & 24 & 4.8 &  & 2.087 & \cite{10_hmxb} \\
  SAX J1802.7$-$2017 & -- & 1.57(25) & 22 & 139.6 &  & 4.570 & \cite{10_hmxb} \\
  OAO 1657$-$415 & -- & 1.74(30) & 17.5 & 37.3 &  & 10.447 & \cite{10_hmxb} \\
  EXO 1722$-$363 & -- & 1.91(45) & 18 & 413.9 &  & 9.741 & \cite{10_hmxb} \\
  4U 1700$-$377 & -- & 1.96(19) & 46 &  &  & 3.412 & \cite{10_hmxb} \\
  J013236.7$+$303228 & -- & 2.0(4) & 11 &  &  & 1.73 & \cite{Bhalerao_J0132367} \\
  Vela X$-$1 & -- & 2.12(16) & 26 & 283.2 &  & 8.964 & \cite{10_hmxb} \\
\hline
  \multicolumn{8}{c}{Binary neutron star mergers} \\
  GW170817 & $1.34^{+0.12}_{-0.09}$ & $1.38^{+0.11}_{-0.11}$ & -- &  &  &  & \cite{gw170817,Zhu_GW170817} \\
  GW190425 & $1.64^{+0.13}_{-0.11}$ & $1.66^{+0.12}_{-0.12}$ & -- &  &  &  & \cite{GW190425,Zhu_GW170817} \\
\hline
  \multicolumn{8}{c}{Neutron star-black hole mergers} \\
  GW191219 & -- & $1.17^{+0.07}_{-0.06}$ & 31.1 &  &  &  & \cite{Abbott_GW191219_GWTC3} \\
  GW200115 & -- & $1.4^{+0.6}_{-0.2}$ & 5.9 &  &  &  & \cite{Abbott_GW200115} \\
  GW200105 & -- & $1.9^{+0.2}_{-0.2}$ & 8.9 &  &  &  & \cite{Abbott_GW200115} \\
  \hline
\hline 
\end{tabular}}
\end{table}

\section{Modelling the accreted masses of recycled pulsars}
\label{sec:model_deltam}
In this section, we describe our methodology for modelling the accreted mass of a recycled pulsar.
Our starting point is the measured properties of recycled pulsars, including the spin period, spin-down rate and mass.
The goal is to estimate the accreted mass gained in the recycling process and hence estimate the birth masses for each of these pulsars.
Our method involves an analytical approach described in the first part of Section \ref{sec:analytic}.
To fully account for the incomplete understanding of the recycling process, we also consider a phenomenological model as detailed in the second part of Section~\ref{sec:phenom}.
As mentioned in Section \ref{sec:data}, we aimed to estimate the birth masses of 53 recycled neutron stars.
They are placed in two groups: group $a$ was composed of all recycled pulsars listed in Extended Data Table \ref{tab:psr_mass} as well as the isolated millisecond pulsar J0030+0451 and PSR J1903+0327; group $b$ consists of 12 ``spider" pulsars, and two recycled neutron stars in low-mass X-ray binaries.
It is generally thought that the pulsars in group $a$ may have experienced notable spin-down since the end of the recycling process, whereas the pulsars in group $b$ probably completed accretion not long before\footnote{ 
The transitional millisecond pulsar J1023+0038 is an excellent example. 
Its optical variability indicates that it had an accretion disk in 2001\cite{Archibald09-J1023,Wang09-J1023disk}.}.

In the standard binary-recycling paradigm, recycled pulsars start their post-recycling evolution on a spin-up line, defined as $\dot{P}\propto P^{4/3}$, in the $P$ versus $\dot{P}$ diagram\cite{GhoshLamb92,tlk12}.
The exact location of such a spin-up line depends on pulsar-specific properties such as mass, moment of inertia and magnetic inclination angle.
For a population of millisecond pulsars, strong observational evidence was recently found for a spin-up line\cite{SpinUp22}.
All pulsars were located to the lower right of this line in the $P$-$\dot{P}$ diagram.
For pulsars in group $a$, we accounted for the pulsar spin-down evolution by speculating where their spin-up lines might lie in the $P$-$\dot{P}$ diagram.
For group $b$, we assumed that the spin-down since the end of the recycling process is negligible and, therefore, that the accreted mass is 
linearly correlated with the \textit{observed} spin period.

The pulsar spin-down evolution can be described in a generic model $\dot{P} \propto P^{2-n}$, where $n$ is the braking index.
If magnetic dipole radiation is solely responsible for pulsar spin down, then $n=3$.
Observationally, the braking index can be determined from $n=2-(P\Ddot{P})/\dot{P}^2$ with $\Ddot{P}$ being the measured second derivative of the spin period.
At present, the braking index has been measured only for a handful of normal pulsars\cite{Archibald16braking}, with measured values ranging from 0.9 to 3.15.
One theory for $n$ deviating from 3 is the magnetic field decay, which could occur for isolated young (normal) pulsars\cite{Goldreich92}.
As magnetic field decay is unlikely for (old) recycled pulsars\cite{AshleyYuri18}, we adopted the standard\footnote{We verified that our results are not sensitive to the adopted value of $n$. For example, results with $n=2$ are essentially the same.} value of $n=3$.

A magnetized neutron star accreting matter from a companion star may reach an equilibrium spin period $P_{\rm eq}$ that is given by the Keplerian orbital period at the inner edge of the accretion disk:
\begin{equation}
\label{eq:Peq}
    P_{\rm eq}=2\pi \sqrt{\frac{r_{\rm{mag}}^{3}}{Gm}}\frac{1}{\omega_{\rm c}} \, ,
\end{equation}
were $G$ is the gravitational constant, $m$ is pulsar mass, and $0.25 < \omega_{\rm c} \leq 1$ is the critical fastness parameter\cite{GhoshLamb92}.
The inner radius of the accretion disk $r_{\rm{mag}} = \phi\, r_{\rm{A}}$, with $r_{\rm{A}}$ being the Alfvén radius, which is a characteristic radius at which magnetic stresses dominate the matter flow in the accretion disk:
\begin{equation}
    r_{\rm{A}} \approx \left(\frac{B^{2}R^{6}}{\dot{M}\sqrt{2Gm}}\right)^{2/7}\, ,
\end{equation}
where $B$ is the magnetic field strength, $R$ is the pulsar radius and $\dot{M}$ is the accretion rate. The parameter $\phi$ is generally assumed to be in the range between 0.5 and 1.4, depending on the poorly known disk-magnetosphere interactions\cite{GhoshLamb92}.

The field strength of a magnetic dipole can be estimated as\cite{spi06}
\begin{equation}
\label{eq:Bfield}
    B=\sqrt{\frac{c^{3} I P \dot{P}}{4\pi^{2}R^{6}}\frac{1}{1+\sin^{2}\alpha}}\, ,
\end{equation}
where $c$ is the speed of light, $I$ is the pulsar moment of inertia and $\alpha$ is the misalignment angle between the magnetic dipole moment and the spin axis.
Note that Equation (\ref{eq:Bfield}) is a convenient approximation of numerical results in ref.\cite{spi06}, which takes into account the plasma contribution in the spin-down torque.
Inspecting Equations (\ref{eq:Peq}-\ref{eq:Bfield}), one finds that $\dot{P}=A P^{4/3}$, with the coefficient $A$ given by\cite{tlk12}
\begin{equation} 
	\label{eq:spin-up-param}
	A = \frac{2^{1/6}G^{5/3}}{\pi^{1/3}c^3}\frac{\dot{M}m^{5/3}}{I}(1+\sin^2\alpha)\phi^{-7/2}\omega_{\rm c}^{7/3}\, .
\end{equation}
As can be seen from Equation (\ref{eq:spin-up-param}), the spin-up line is not uniquely defined. 
Instead, the coefficient $A$ depends on pulsar-specific parameters.
In this work, we adopt $A=A_{\rm max}=\unit[3.76 \times 10^{-15}]{s}^{-4/3}$ as the upper limit for the population of recycled pulsars\cite{SpinUp22}.
We also adopted the following empirical relation for the moment of inertia\cite{ls05}
\begin{equation} 
	\label{eqn:inertia}
	 I \approx 0.237 mR^2\Bigg[1+4.2\frac{m}{\rm{M}_\odot}\frac{\rm km}{R} + 90\Bigg(\frac{m}{\rm{M}_\odot}\frac{\rm [km]}{R} \Bigg)^4 \Bigg] \, .
\end{equation}

The intersection between the spin-down line $\dot{P} \propto P^{2-n}$ and the limiting spin-up line (with $A=A_{\rm max}$) gives a minimum initial spin period for recycled pulsars.
However, for some pulsars this minimum period (determined from the intersection of two lines) was unphysically small, in which case we set the minimum spin period to $\unit[1]{ms}$.
This is close to the break-up spin period of neutron stars\cite{KepFreq_NS09}. Also note that in a recent numerical simulation study, the spin period of a neutron star at the end of the recycling process was found to be $\gtrsim \unit[1]{ms}$ for a range of binary evolution calculations\cite{LI2021}.
The time spent for a pulsar to evolve from an
initial spin period ($P_0$) to its current location in the $P$-$\dot{P}$ diagram is 
\begin{equation}
\tau=\frac{P}{(n-1)\dot{P}}\bigg[1-\bigg(\frac{P_0}{P}\bigg)^{n-1}\bigg]\, .
\end{equation}
Here if $P_0 \ll P$ and $n=3$, this reduces to the expression for the characteristic age ($\tau_{\rm c}$).
We denote the time for a pulsar to evolve from either the spin-up line or $P_{0}=\unit[1]{ms}$ to its current period as the maximum age $\tau_{\rm{max}}$. If $\tau_{\rm{max}}$ is greater than the age of Milky Way, $\unit[11]{Gyr}$ (refs.\cite{2012Natur.486...90K,2011A&A...533A..59J}), we set it to $\unit[11]{Gyr}$.

In our calculations, we first obtained pulsar intrinsic spin-down rate $\dot{P}_{\rm{int}}$ by correcting the measured $\dot{P}$ with contribution from the Shklovskii effect\cite{1970shk} and Galactic potential\cite{1991Damour,2018MNRAS.478.2359L} using the package {\tt{GalDynPsr}}\cite{2018ApJ...868..123P}.
For three pulsars in globular clusters, PSRs J1910$-$5959A, B2127$+$11C and J1807$-$2500B, the contribution from cluster potential was also included\cite{2012ApJ...745..109L}.
%Extended Data Table \ref{tab:psr_birth} lists the intrinsic spin-down rate $\dot{P}_{\rm{int}}$, characteristic age, minimum spin period $P_{\rm lim}$ and its corresponding initial spin-down rate $\dot{P}_{\rm lim}$ assuming $n=3$, and maximum age $\tau_{\rm{max}}$.

\begin{figure}
\centering
\includegraphics[width=120mm]{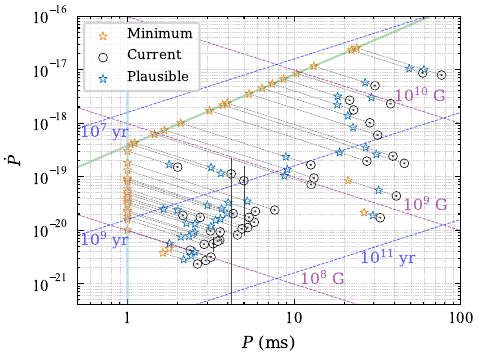}\\
\caption{\textbf{The pulsar $P$-$\dot{P}$ diagram for 39 recycled pulsars}. Black circles mark their current locations (where the error bars on the intrinsic $\dot{P}$ are too small to be seen except for the two globular-cluster pulsars), whereas orange and blue stars mark the minimum and plausible initial spin periods (assuming a breaking index $n=3$), respectively. The solid green line is the limiting spin-up line $\dot{P} \propto P^{4/3}$ inferred for the population of millisecond pulsars\cite{SpinUp22}.}
\label{fig:ppdot}
\end{figure}

Extended Data Figure \ref{fig:ppdot} shows the locations of 39 recycled pulsars (in group $a$) in the $P$-$\dot{P}$ diagram.
Their observed locations with intrinsic $\dot{P}_{\rm{int}}$ are marked by black circles.
Orange stars indicate the minimum spin period $P_{\rm lim}$ and its corresponding initial spin-down rate $\dot{P}_{\rm lim}$.
For a braking index of 3, black dotted lines ($\dot{P} \propto P^{-1}$) connect the black circles and orange stars.
Note that $P_{\rm lim}$ and $\dot{P}_{\rm lim}$ are determined by three constraints: the limiting spin-up line (green line in Extended Data Figure \ref{fig:ppdot}), a minimum spin period of $\unit[1]{ms}$ (a vertical blue line), and a maximum age of $\unit[11]{Gyr}$.
For orange stars that do not fall onto the green or blue lines, their locations are a result of the age limit.
Light blue stars in this plot indicate plausible places of recycled pulsars at the end of recycling process. 
As we assumed that the age of a recycled pulsar is uniformly distributed in $[0, \tau_{\rm{max}}]$, blue stars correspond to an age of $\tau_{\rm{max}}/2$.

\subsection{An analytical model}
\label{sec:analytic}

In our analytical model, we approximate the equilibrium spin period of a neutron star at the end of the recycling process as the minimum spin period accelerated from rest\cite{Lipunov84}:
\begin{equation}
\label{eq:PeqDm}
    P_{\rm eq}=\frac{3 \pi I}{\sqrt{G r_{\rm mag}}} \left(m^{3/2} - m^{3/2}_{\rm ini} \right)^{-1} \, ,
\end{equation}
where $m_{\rm ini}=m-\Delta m$ is the neutron-star mass before accretion (its birth mass), $m$ is the measured neutron-star mass and $\Delta m$ is the accreted mass that is being modelled here.

From Equations (\ref{eq:Peq})-(\ref{eqn:inertia}) and (\ref{eq:PeqDm}), it can be seen that, the spin-up line and the accreted mass is uniquely determined for a given set of parameters $\{\dot{M}, \alpha, R, \phi, \omega_{\rm c} \}$, if we assume that a recycled pulsar starts its rejuvenated life from the spin-up line and spins down due to magnetic dipole braking.
Therefore, we modelled $\Delta m$ in a parameter-agnostic way.
For each recycled pulsar, we set the following priors on its recycling parameters.
The accretion rate, when expressed in units of the Eddington accretion rate, follows a log-uniform distribution between 0.01 and 3. 
The upper end was chosen to allow for possible super-Eddington accretion\cite{faa+11}.
The magnetic inclination angle $\alpha$ is distributed isotropically, with $\cos \alpha$ uniform between $-1$ and 1.
The dimensionless parameters $\phi$ and $\omega_{\rm c}$ are uniformly distributed in the ranges $[0.5,1.4]$ and $[0.25,1]$, respectively.
The pulsar radius follows a Gaussian distribution with a mean of $\unit[11.9]{km}$ and a standard deviation of $\unit[0.85]{km}$, based on gravitational-wave observations of the first neutron star merger GW170817\cite{GW170817eos}.
We have verified that the effect of using a mass-radius relation for a specific equation of state that is consistent with current observational constraints, for example, the AP4 equation of state\cite{Lattimer01eos}, is of the order of $\lesssim 10\%$.

For each pulsar, we generated a sample in the six-dimensional parameter space 
$\{m, \dot{M}, \alpha, 
\newline R, \phi, \omega_{\rm c} \}$, and draw a spin-up line (which must lie to the upper left of the current location of pulsar in the $P$-$\dot{P}$ diagram).
Then, the intersection between this spin-up line and the constant magnetic field strength line $\dot{P} \propto P^{-1}$ determines an equilibrium spin period $P_{\rm eq}$.
The accreted mass $\Delta m$ was then computed from Equation (\ref{eq:PeqDm}).
Note that the statistical uncertainties of $P$ and $\dot{P}$ were ignored, which is an excellent approximation for all pulsars except one globular-cluster pulsar.
%; 2) in deriving $P_{\rm eq}$, three constraints are applied as illustrated in Figure \ref{fig:ppdot}.

For recycled neutron stars in group $b$, we adopted a simple scaling relation (plotted as the red dashed line in Extended Data Figure~\ref{fig:DeltaM_spin}) between the accreted mass $\Delta m$ and the observed spin period: $\Delta m /\mathrm{M}_{\odot} = 0.2 P_{\rm ms}^{-1}$ where $P_{\rm ms}$ is the spin period in milliseconds; see, for example, equation (6) in ref.\cite{Kremer22msp}.
Such a relation can be obtained from Equation (\ref{eq:PeqDm}) when $\Delta m \ll m$.

\subsection{A phenomenological model}
\label{sec:phenom}

%\subsubsection{The initial spin periods of recycled pulsars}
%\label{sec:Pini}

Here we estimate the accreted masses of recycled neutron stars using a phenomenological relation between $\Delta m$ and the initial spin periods $P_{\rm{ini}}$.
As in the analytic model, for group $b$ neutron stars, we assumed that the observed spin periods are their initial spin periods.
For group $a$ pulsars, the initial spin period was determined as follows.
We assumed that the true age of a recycled pulsar is uniformly drawn from $[0, \tau_{\rm{max}}]$.
For a given true age, we evolvd the pulsar back in time following the spin-down equation $\dot{P} \propto P^{2-n}$ to find its $P_{\rm{ini}}$.
We generated $10^{4}$ random ages uniformly drawn from $[0, \tau_{\rm{max}}]$ and obtained the distribution of initial spin periods $P_{\rm{ini}}$.
%The mean initial periods $\overline{P_{\rm{ini}}}$ are listed in Extended Data Table \ref{tab:psr_birth}, and also marked as blue stars in Figure \ref{fig:ppdot}.
For a given $P_{\rm{ini}}$, $\Delta m$ is drawn from a Gaussian distribution with a mean of $\mu_{M}/\mathrm{M}_{\odot} = 0.3 P_{\rm ms}^{-1/2}$ (with $P_{\rm ms}$ being the initial spin period measured in milliseconds) and a standard deviation of $0.2\mu_{M}$.
The mean in this recipe is motivated by numerical simulations that take into account the neutron star spin-down due to propeller effects during the recycling process\cite{LI2021}; see the blue dots Extended Data Figure \ref{fig:DeltaM_spin}.
The $20\%$ dispersion is the typical spread found in the analytical prescription.

%\subsubsection{Estimating the accreted masses of recycled pulsars}
%\label{sec:DeltaM}

%In the recycling process, the amount of mass accreted by a neutron star depends on the mass-transfer timescale, accretion efficiency, magnetic field strength, among other factors.
%Under certain simplifications and considering only the spin-up process, it was shown that $\Delta m$ scales as $P_{\rm ini}^{-4/3}$\cite{Alpar82Nat,tlk12}, or similarly $\Delta m \propto P_{\rm ini}^{-1}$ if $\Delta m \ll m$; see Equation (\ref{eq:PeqDm}).
%However, as pointed out by ref.\cite{tlk12}, the $\Delta m \propto P_{\rm ini}^{-4/3}$ relation only gives a lower limit on $\Delta m$ because of, for example, the propeller effect\cite{IllaSunya75} and accretion disc instabilities\cite{Pringle81disc}.

\begin{figure}
\centering
\includegraphics[width=115mm]{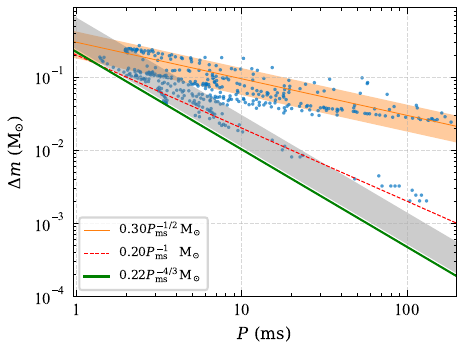}\\
\caption{\textbf{The accreted mass-spin period ($\Delta m$-$P$) correlation used to correct for mass accreted by recycled pulsars}. The green solid line depicts a lower limit on the accreted mass\cite{Alpar82Nat,tlk12}, while the orange solid lines and shaded band represent the mean and 90\% credible region of our phenomenological model. The red dashed line is a simple scaling used in the literature\cite{Lipunov84,Kremer22msp}. The grey shaded band encompasses 90\% credible region of our analytical model. Blue dots are from numerical simulations of the recycling process performed in ref.\cite{LI2021}.}
\label{fig:DeltaM_spin}
\end{figure}

%In this work, we consider two prescriptions for the accreted mass-spin period correlation.
%The ``low"-accreted mass prescription is given by $\Delta m_{\rm min}/\mathrm{M}_{\odot} = 0.22 P_{\rm ms}^{-4/3}$, where $P_{\rm ms}$ is the spin period in ms.
%In comparison to Equation (14) in ref.\cite{tlk12}, we ignore the weak dependency ($\sim M^{1/3}$) on the neutron star mass.
%In the ``high"-accreted mass prescription, we add a level of stochasticity to $\Delta m$ to account for uncertainties in the recycling process.

Extended Data Figure \ref{fig:DeltaM_spin} illustrates our prescriptions for the accreted mass-spin period correlation.
The green and orange solid lines represent $\Delta m_{\rm min}/\mathrm{M}_{\odot} = 0.22 P_{\rm ms}^{-4/3}$ and $\mu_{M}/\mathrm{M}_{\odot} = 0.3 P_{\rm ms}^{-1/2}$, respectively.
The grey- and orange-shaded bands encompass the 90\% credible regions of the analytical and phenomenological models, respectively.
The red dashed line is $\Delta m /\mathrm{M}_{\odot} = 0.2 P_{\rm ms}^{-1}$.
Results from numerical simulations of the recycling process in ref.\cite{LI2021} are marked as blue dots. Note that their calculations were performed for a fiducial value of the initial magnetic field strength and neutron star moment of inertia.
Our models captures the overall trend and scatter of the numerical results well.
The difference between grey and orange bands for $P \gtrsim \unit[10]{ms}$ is exaggerated due to the logarithmic scale of the vertical coordinate in Extended Data Figure \ref{fig:DeltaM_spin}.
%The mean accreted masses for 39 recycled pulsars using the analytical and ``high"-accreted mass recipes are listed in Extended Data Table \ref{tab:psr_birth}.

\begin{figure}
\centering
\includegraphics[width=120mm]{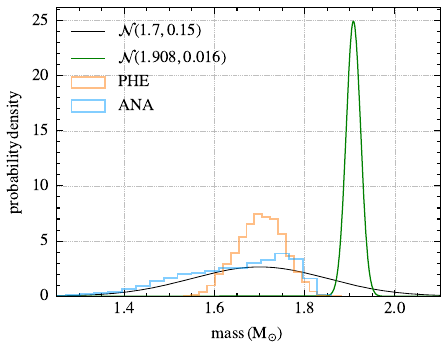}\\
\caption{\textbf{The mass probability distribution of PSR J1614$-$2230}. The green curve depicts the measured mass\cite{NANOGrav11yr}, whereas the black curve is the birth mass deduced from detailed binary evolution calculations\cite{Tauris1614mass}, based on the original measured mass of $\unit[1.97 \pm 0.04]{\rm M_{\odot}}$\cite{Demorest10Nat}. Birth masses estimated from our analytical and phenomenological models are shown in blue and orange, respectively.}
\label{fig:1614_mass}
\end{figure}

\subsection{A case study of PSR J1614$-$2230.}

PSR J1614$-$2230, which has an observed spin period of $\unit[3.15]{ms}$, was the first binary pulsar found to have a measured neutron-star mass around $\unit[2]{\rm M_{\odot}}$.
The pulsar mass was originally measured as $\unit[1.97 \pm 0.04]{\rm M_{\odot}}$, and the neutron star is in an 8.7-day orbit with a $\unit[0.5]{\rm M_{\odot}}$ white dwarf companion\cite{Demorest10Nat}.
By performing detailed binary evolution calculations, Tauris et al.\cite{Tauris1614mass} suggested that the most probable progenitor system of PSR J1614$-$2230 is a $\unit[1.7 \pm 0.15]{\rm M_{\odot}}$ neutron star going through ``case A'' Roche lobe overflow (and, thus, accreting mass) from an $\sim\unit[4.5]{\rm M_{\odot}}$ donor star.
An independent binary evolution study found a \textit{minimum} initial (birth) mass of $\unit[1.6 \pm 0.1]{\rm M_{\odot}}$ (ref.\cite{Lin1614mass}).

The mass measurement of PSR J1614$-$2230 was updated in 2018 to a slightly lower value of $\unit[1.908 \pm 0.016]{\rm M_{\odot}}$ with extended observations\cite{NANOGrav11yr}.
From our accreted-mass modelling, we found that its birth mass was around $\unit[1.7]{\rm M_{\odot}}$, in excellent agreement with binary evolution calculations.
Extended Data Figure \ref{fig:1614_mass} shows the measured and inferred birth-mass distributions of PSR J1614$-$2230.
The blue and orange curves indicate the birth masses estimated from our analytical and phenomenological modelling, respectively.

\section{Bayesian inference}
\label{sec:bayes}

In this section, we briefly describe the method of Bayesian inference used to determine the birth-mass function of neutron stars; a more general introduction can be found in refs.\cite{2019MNRAS.486.1086M,2019PASA...36...10T}.

The data analyzed are posterior distributions for 90 neutron stars.
The original dataset, denoted as OBS, consists of Gaussian mass posteriors used to approximate measurements reported in the literature (as summarized in Extended Data Tables \ref{tab:psr_mass} and \ref{tab:psr_mass1}) or posterior samples, if available.
For 53 of the recycled pulsars in our compilation, we also applied probabilistic accreted mass corrections to obtain posteriors for their birth masses.
The dataset obtained with the analytical (phenomenological) prescription described in Section \ref{sec:analytic} is denoted as ANA (PHE).
A dataset that includes only those measurements of slow (non-recycled) neutron stars is denoted as SLOW.
For comparison, the dataset that includes only observed masses of 53 recycled pulsars is denoted as REC.
The $\rm{ANA}_{m_r}$ ($\rm{PHE}_{m_r}$) dataset includes the birth masses of 53 recycled pulsars only after applying the analytical (phenomenological) accreted-mass correction.
The ANA$_{\rm {radio}}$ dataset is a subset of ANA that includes only mass measurements obtained by the radio pulsar timing method (i.e., all in Extended Data Table \ref{tab:psr_mass}, plus PSR J1903+0327).
%The naming for other data sets obtained with the phenomenological prescriptions (Section \ref{sec:phenom}) works as follows.
%The low- and high-accreted mass prescription is paired with the slow- and fast-spin scenarios, leading to 4 data sets: LS, LF, HS, HF.
%In addition, we generate LF2 and HF2 assuming a braking index of $n=2$ for pulsar spin-down evolution, while all the other data sets have $n=3$.
%Individual mass posteriors are shown in Figure \ref{fig:mass_post} for the OBS, ANA and HF data sets for 39 recycled pulsars.

Let $p(m^i | d^i)$ be the mass posterior distribution of the $i$th neutron star, given some observations $d^i$.
Our goal is to find a model that can best describe the collection of mass posteriors for $N$ sources.
We introduce a conditional prior for neutron-star mass $\pi(m^{i} | \Lambda)$ for a given model $M$ that includes hyperparameters $\Lambda$.
The hyperposterior distribution of $\Lambda$ can be written as
\begin{equation}
\label{eqn:hyperpost}
    p(\Lambda|\{d\},M) = 
\frac{1}{{\cal Z}_\Lambda (M)}
\prod_{i=1}^N \int {\rm d}m^i \,{\cal L}(d^i|m^i) \pi(m^i|\Lambda) \pi(\Lambda |M)\ ,
\end{equation}
where $\{d\}$ denotes the original observations that provide individual neutron-star mass measurements, $\pi (\Lambda |M)$ is the prior distribution for $\Lambda$ and the term ${\cal L}(d^i|m^i)$ is the likelihood function of the data given $m^i$, which is related to the mass posterior through
\begin{equation}\label{eqn:mlikelihood}
{\cal L}(d^i|m^i) = \frac{{\cal Z}_0^i}{\pi_0(m^i)} p(m^i|d^i)\ ,
\end{equation}
where $\pi_0(m^{i})$ is the initial prior used to derive the mass posterior (assumed to be flat here) and
${\cal Z}_0^i$ is the initial evidence, which is a factor that will ultimately cancel out through either normalization of the hyperposterior or the construction of a Bayesian evidence ratio.

The term ${\cal Z}_\Lambda (M)$ in Equation (\ref{eqn:hyperpost}) is the hyperevidence
\begin{equation}
\label{eq:ZLambda}
{\cal Z}_\Lambda (M) = \int {\rm d}\Lambda \,
\prod_{i=1}^N \int {\rm d}m^i {\cal L}(d^i|m^i)
\pi(m^i|\Lambda)  \pi(\Lambda |M)\, .
\end{equation}
In practice, we can replace the integral over ${\rm d} m^i$ with a summation over the conditional prior at posterior samples of $m^i$, so that the hyperevidence is given by
\begin{equation}
\label{eq:ZLambda_samp}
{\cal Z}_\Lambda (M) = \int {\rm d}\Lambda \,
\prod_{i=1}^N \frac{{\cal Z}_0^i}{n_i}\sum^{n_i}_{k=1} \frac{\pi(m^i_k|\Lambda)}{\pi_{0}(m^i_k)}  \pi(\Lambda |M)\, ,
\end{equation}
where $n_i$ is the number of mass posterior samples for the $i$th neutron star.

To find the model that best fits the birth-mass function of neutron stars, we compute the Bayes factor between two models ($M_1$ and $M_2$), defined as follows
\begin{equation}
\label{eq:BF}
{\rm BF}_{M_2}^{M_1} = \frac{{\cal Z}_{\Lambda}(M_1)}{{\cal Z}_{\Lambda}(M_2)}\ .
\end{equation}
We followed the interpretation of Bayes factors outlined by ref.\cite{Raftery95} and choose a threshold for BF of 150, corresponding to a natural logarithmic BF of 5, as required for confident model selection.
Note that the Bayes factor is equal to the odds, which is defined as the ratio of the posterior probabilities between two models, assuming that their prior probabilities are equal.

The likelihood function adopted above is applicable when individual neutron-star mass measurements are independent.
For Galactic double neutron star systems, the mass measurements are correlated because of the significant constraint on the total mass of two neutron stars, as demonstrated in figure 1 in ref.\cite{farrow19dns}.
In this case, the joint posteriors of two masses should be used following the formalism outlined in Section 3.1 of ref.\cite{farrow19dns}.
To illustrate this point explicitly, the total likelihood that is hidden in Equation (\ref{eqn:hyperpost}) is
\begin{equation}
\label{eq:total_like}
{\cal L} (\{d\} | \Lambda) =
\prod_{i=1}^{90} \int {\rm d}m^i \,{\cal L}(d^i|m^i) \pi(m^i|\Lambda)  \, ,
\end{equation}
which should be modified as follows:
\begin{equation}
\label{eq:total_like_different_dataset}
\prod_{i=1}^{64} \int {\rm d}m^i \,{\cal L}(d^i|m^i) \pi(m^i|\Lambda) \times \prod_{j=1}^{13} \int {\rm d}m_{1}^j {\rm d}m_{2}^j \,{\cal L}(d^j | m_{1}^j,m_{2}^j) \pi(m_{1}^j|\Lambda) \pi(m_{2}^j|\Lambda)  \, .
\end{equation}
We divided the 90 neutron stars into two groups: the first group includes 64 neutron stars whose mass posteriors are truly independent; the second group contains 13 pairs of neutron stars (labelled as $m_{1}^j$ and $m_{2}^j$) that are in double neutron star systems.
For the second group, the individual likelihood function ${\cal L}(d^j | m_{1}^j,m_{2}^j)$ is proportional to the joint posterior $p(m_{1}^j, m_{2}^j | d^j)$.

In our analysis, we used the software package {\tt BILBY} (ref.\cite{ashton2019bilby}) for parameter estimation and evidence calculation and the ${\tt{gwpopulation}}$ package \cite{2019PhRvD.100d3030T} for population modelling.

\section{Parametric models for the neutron star mass function}
\label{sec:model}

In this section, we describe a range of parametric models for $\pi(m | \Lambda)$ and specify the priors adopted for hyperparameters $\pi (\Lambda |M)$.

Our base model is a uniform distribution:
\begin{align}
\pi(m \mid {\Lambda})&=\pi \left(m \mid\left\{m_{\min }, m_{\max }\right\}\right) \nonumber \\
&=\begin{cases} \frac{1}{m_{\max} - m_{\min}} & m_{\min } \leq m \leq m_{\max } \\ 0 & \text { otherwise }\end{cases},
\end{align}
where $m_{\min}$ and $m_{\max }$ are the maximum and minimum neutron-star mass, respectively.
For the uniform model and all other models where applicable, we adopted uniform priors in the ranges $[0.9, 2.9] \mathrm{M}_{\odot}$ and $[1.9, 2.9] \mathrm{M}_{\odot}$ for $m_{\min}$ and $m_{\max }$, respectively.
The prior edges were chosen to be the same as in ref.\cite{alsing2018} to facilitate comparisons.

The probability distribution of a Gaussian model is given by
\begin{align}
\pi(m \mid \Lambda)&=\pi(m \mid\{\mu, \sigma\}) \nonumber \\
&=\frac{1}{\sigma \sqrt{2 \pi}} \exp \left[-\frac{1}{2}\left(\frac{m-\mu}{\sigma}\right)^{2}\right],
\end{align}
where $\mu$ and $\sigma$ are the mean and standard deviation, respectively.
This model can be extended to multi-component Gaussians.
As in ref.\cite{alsing2018}, we also considered the two-Gaussian (2G) and three-Gaussian (3G) models.
For 2G, we have $\Lambda = \{\mu_{1}, \sigma_{1}, \mu_{2}, \sigma_{2}, \alpha \}$, whereas for 3G model, $\Lambda = \{\mu_{1}, \sigma_{1}, \mu_{2}, \sigma_{2},\mu_{3}, \sigma_{3}, \alpha, \beta \}$.
Here $\alpha$ and $\beta$ are the weights of the first and second Gaussian components, respectively.
We adopt uniform priors in the ranges $[0.9, 2.9] \mathrm{M}_{\odot}$ and $[0.01, 2] \mathrm{M}_{\odot}$ for the mean and standard deviation, respectively, with a further constraint of $\mu_i < \mu_{i+1}$ for multi-component Gaussian models.
The priors for weight parameters $\alpha$ and $\beta$ are uniform between 0.01 and 1.
For 2G, we also added a minimum-mass ($m_{\min}$) cutoff, a maximum-mass ($m_{\max}$) cutoff or both.
These models are denoted as 2G$_{\rm min}$, 2G$^{\rm max}$, and 2G$\rm {^{max}_{min}}$, respectively.
In the case with a mass cutoff (truncated Gaussians), the probability distribution is properly normalized.

The probability distribution in the log-Uniform model is
\begin{align}
\pi(m \mid {\Lambda})&=\pi \left(m \mid\left\{m_{\min }, m_{\max }\right\}\right) \nonumber \\
&=\begin{cases} \frac{1}{m \log(m_{\max} /m_{\min})} & m_{\min } \leq m \leq m_{\max } \\ 0 & \text { otherwise }\end{cases},
\end{align}
Whereas the log-Normal distribution is defined as
\begin{align}
\pi(m \mid {\Lambda})&=\pi(m \mid\{\mu, \sigma\}) \nonumber \\
&=\frac{1}{\sqrt{2 \pi} m \sigma} \exp \left[-\frac{(\log m-\mu)^{2}}{2 \sigma^{2}}\right].
\end{align}
The prior edges of the log-Normal model are chosen to match those of the Gaussian model.

In the power-law model, the neutron star mass function is given by
\begin{align}
\pi(m \mid {\Lambda})&=\pi \left(m \mid\left\{m_{\min }, m_{\max }, \alpha \right\}\right) \nonumber \\
&=\begin{cases}\frac{1-\alpha}{m_{\max }^{1-\alpha}-m_{\min }^{1-\alpha}} m^{-\alpha} & m_{\min } \leq m \leq m_{\max } \\ 0 & \text { otherwise }\end{cases}.
\end{align}
Here the power-law index takes a uniform prior between $-5$ and 25.
The power-law distribution can be modified to allow a smooth turn-on over a range of $\delta m$, which takes a uniform prior between 0.01 and $1 \mathrm{M}_{\odot}$ in our analysis.
Such a TOP model is defined as
\begin{align}
\pi(m \mid {\Lambda})&=\pi \left(m \mid\left\{m_{\min }, m_{\max }, \alpha, \delta m \right\}\right) \nonumber \\
&= m^{-\alpha} S\left(m, m_{\min }, \delta m\right) \mathcal{H}\left(m_{\max }-m\right),
\end{align}
where $\mathcal{H}$ is the Heaviside step
function and $S$ denotes a smoothing function, which is designed to rise from 0 at $m_{\min}$ to 1 at $(m_{\min}+\delta m)$. In the interval of $m_{\min} < m < m_{\min}+\delta m$, the function $S$ can be expressed by
\begin{equation}
S\left(m, m_{\min }, \delta m\right) =\left[\exp \left( \frac{\delta m}{m-m_{\min}}+\frac{\delta m}{m-m_{\min}-\delta m}\right)+1\right]^{-1}\ . 
\end{equation}
Note that in plots of the posterior distributions of TOP parameters, we show the parameter $m_{\rm peak}$ which defines the peak location of the distribution.

The gamma distribution is defined as
\begin{equation}
\pi(m \mid {\Lambda}) = \pi (m \mid \alpha,\theta)=\frac{m^{(\alpha-1)} \exp{\left(-\frac{m}{\theta}\right)}}{\theta^{\alpha} \Gamma(\alpha)}\ ,
\end{equation}
where $\Gamma(\alpha)$ is the gamma function, the $\alpha$ is the shape parameter and $\theta$ is the scalar parameter. We adopted uniform priors for $\alpha$ and $\theta$ in the range of $(0,80)$ and $(0.01,0.1)$, respectively.

The skewed Student's $t$-distribution (SST) can be parameterized as\cite{sst_jones2003,sst_wurtz2006}
\begin{equation}
\label{eqn:sst}
\pi(m \mid {\Lambda}) = \pi (m \mid \mu, \sigma, \nu, \tau)= \begin{cases}\frac{c}{\sigma_{0}}\left[1+\frac{\nu^{2} z^{2}}{\tau}\right]^{-(\tau+1) / 2} & \text { if } m < \mu_{0}, \\ \frac{c}{\sigma_{0}}\left[1+\frac{z^{2}}{\nu^{2} \tau}\right]^{-(\tau+1) / 2} & \text { if } m \geq \mu_{0}.\end{cases}
\end{equation}
Here $\mu$ and $\sigma$ are the mean and standard deviation of the Gaussian distribution, respectively. The parameters $\nu$ and $\tau$ are used to characterize the degree of skewness.
In Equation (\ref{eqn:sst}), $\mu_{0}=$ $\mu-\sigma \kappa / s$, 
$\sigma_{0}=\sigma / s$,
$z=\left(m-\mu_{0}\right) / \sigma_{0}$, 
and $c=2 \nu\left[\left(1+\nu^{2}\right) B(1 / 2, \tau / 2) \tau^{1 / 2}\right]^{-1}$ with $\kappa$ and $s$ given by $\kappa=2 \tau^{1 / 2}\left(\nu-\nu^{-1}\right)[(\tau-1) B(1 / 2, \tau / 2)]^{-1}$ and $s=\left[ \frac{\tau}{(\tau-2)}\left(\nu^{2}+\nu^{-2}-1\right)-\kappa^{2} \right]^{1/2}$, respectively; $B$ is the beta function. Priors for $\mu$ and $\sigma$ were the same as those of the Gaussian-family models, and we adopted uniform priors for $\nu$ and $\tau$ in the ranges $(0, 8) $ and $(2, 20)$, respectively.

\section{Further details of analysis results and discussions}
\label{sec:result}

Extended Data Table \ref{tab:BF_matrix1} lists the log Bayes factors for 15 models in comparison to the uniform distribution for 9 data sets, including the original observed masses (OBS), the observed masses of double neutron star systems (OBS$_{\rm {DNS}}$), and accounting for various accreted-mass corrections.
One can see that the turn-on power law (TOP) distribution is the best-performing model across 9 different data sets.
The TOP model is favoured against the two-Gaussian (2G) model with a natural logrithmic Bayes factor greater than 5 for both the analytical and phenomenological prescriptions of accreted-mass corrections.
We consider two additional features in the mass spectrum on top of the TOP model: a maximum-mass cutoff (``$\mathrm{TOP}^{\max }$") and a Gaussian peak (denoted as ``TOPG").
We find that there is marginal support for a maximum-mass cutoff but no support for a Gaussian peak.
For simplicity, in referring to the preferred model from data, we do not distinguish between TOP and $\mathrm{TOP}^{\max }$.

In order to verify the significance of our results, we simulate a sample of 87 neutron star mass measurements.
The mean values of these neutron star masses are drawn from a 2G distribution, with parameters corresponding to the maximum a posteriori probability values inferred from the ANA data set; their measurement uncertainties are the same as the actual data set; here we rank the mean values of both simulated and real neutron star masses, and assign the measurement uncertainty of actual neutron star mass measurement to its simulated counterpart.
We generate 200 such synthetic data sets, and compute the Bayes factor between a TOP and 2G model for each of them. Extended Data
 Figure \ref{fig:BF_hist} shows the distribution of the log Bayes factor obtained from such a simulation.
The red dashed line marks the log Bayes factor of 5.8 computed for the actual ANA data set.
The false alarm probability of this Bayes factor is found to be 2.5\%.

Now we get back to Extended Data Table \ref{tab:BF_matrix1} and provide some further interpretations into the results.
%From Figure \ref{fig:BF_hist}, we can see that a log Bayes factor of 5 indeed implies very strong support for one model against another, with a false alarm probability of few percent.
First, we find that one can obtain a better fit of the neutron star mass function by adding a minimum- or maximum-mass cutoff or both to the 2G model.
For example, the 2G$^{\rm max}$ model is favoured with a modest log Bayes factor of 1.1 against the 2G model for the ANA data set.
This is in agreement with ref.\cite{alsing2018}, where a log Bayes factor of 1.55 was found (albeit using a different set of neutron star mass measurements).
In our analysis, again for the ANA data set, the model 2G$\rm {^{max}_{min}}$ improves upon 2G$^{\rm max}$ with a log Bayes factor of 2.9.
However, such a complex model with 7 free parameters is still disfavoured by a log Bayes factor of 1.8, in comparison to the TOP model.
Therefore, we conclude that 1) the TOP model provides a simple and compelling fit to the neutron star mass function, no matter it is the observed masses or birth masses; 2) it is possible that future neutron star mass measurements will reveal more features in the mass spectrum, however we find no evidence of such features in current observations.
The posterior median and 1-$\sigma$ credible intervals of TOP parameters, along with their prior ranges are given in Extended Data Table \ref{tab:prior_posterior}, to facilitate the usage of our model in future neutron star population studies.

In further comparison to previous studies, we note that the mass distribution of slow neutron stars in double neutron star systems was found to be uniform in ref.\cite{farrow19dns}, whereas the uniform model is disfavoured in our analysis for the SLOW dataset (see Extended Data Table \ref{tab:BF_matrix1}). We first confirmed that we can reproduce the results of ref.\cite{farrow19dns} using the same data and comparing the same set of models. The discrepancy was largely due to a small number of measurements (12 versus 34). Additionally, the results of ref.\cite{farrow19dns} were also likely biased because of the incorrect assumption that the binary orbits of some double neutron star systems are randomly oriented\footnote{For this reason, we exclude in this work pulsar mass measurements that are dependent on the unknown binary orbital inclination angle.}; a majority of binary pulsars in Extended Data Table \ref{tab:psr_mass} have measured inclination angles close to $90^{\circ}$ (i.e., edge-on), which is the preferred orbital orientation for the pulsar mass to be measurable via Shapiro delays.

Extended Data Figure \ref{fig:TOP_corner} shows the posterior distribution of model parameters of the TOP model for the OBS (in orange), ANA (in blue), and PHE (in black) data set.
Comparing three cases, we find that although there are noticeable shifts, the three sets of posteriors are consistent with each other within uncertainties.
This is because the accreted-mass corrections are comparable or smaller than the measurement uncertainties of original data.
This plot also explains why different accreted-mass prescriptions lead to similar model selection results in Extended Data Table \ref{tab:BF_matrix1}.
%For the SLOW data set, it is interesting to note that whereas posteriors of $m_{\rm min}$ and $\delta m$ are in good agreement with the other two cases, the posterior distribution of $\alpha$ shifts to significantly higher values.
%Suppose the true mass function is given by the TOP model with parameters inferred from the HF data set, the above feature with SLOW is unsurprising.
%For a small number of observations, it is less likely to see high-mass neutron stars, leading to greater posterior support at higher power-law index $\alpha$.
%Overall, the posteriors in $\alpha$ are \textit{broadly} consistent across three cases shown in Figure \ref{fig:TOP_corner}.

Extended Data Figure \ref{fig:TOP_corner1} shows the comparison of posterior distributions of the TOP model for different subsets of data.
In the top panel, we demonstrate how the correction of accreted mass of recycled pulsars bridges the gap between two distinct subclasses -- slow (SLOW) and recycled (REC) neutron stars.
Although the uncertainties are large and therefore we cannot make conclusive statements, one can see that masses of recycled neutron stars are on average greater than those of slow neutron stars, and that our scheme accreted-mass corrections is effective because the ANA$_{m_{r}}$ posterior distribution lies in between those of SLOW and REC.
To test how our results are sensitive to the selection of neutron mass measurements obtained using different methods, we restrict our analyses to data from radio pulsar timing observations (denoted as ANA$_{\rm {radio}}$) and find that the TOP model is favoured against other models with similar significance (with respect to the full data set).
In the bottom panel of Extended Data Figure \ref{fig:TOP_corner1}, we show that the posterior distributions of the TOP model for the radio-only data set agree well with results of the full data set.
The power-law index parameter $\alpha$ is poorly measured for ANA$_{\rm {radio}}$ due to smaller number of measurements.

\clearpage
%\clearpage

\begin{table*}
\begin{center}
 \caption{\textbf{The natural logarithm of Bayes factor between a listed model and a uniform distribution for different data sets.} Model names are: TOP -- turn-on power law, G -- Gaussian distribution, TOPG -- turn-on power law plus a Gaussian peak, 2G and 3G indicates two- and three-component Gaussian distribution, respectively, SST -- skewed Student's t distribution, logN -- log normal distribution, logU -- log uniform distribution, Gamma -- gamma distribution. The presence of a superscript or subscript indicates a maximum- or minimum-mass cutoff applied to the continuous distribution.}
 \label{tab:BF_matrix1}
\begin{tabular}{l|ccccccccc}
\hline \hline
  & $\rm{OBS}$ & $\rm{ANA}$ & $\rm{PHE}$ & $\rm{ANA}_{m_r}$  & $\rm{PHE}_{m_r}$ & REC & SLOW & ANA$_{\rm {radio}}$ & OBS$_{\rm {DNS}}$ \\
\hline $\mathrm{TOP}^{\max }$ & 13.2 & 16.1 & 13.1 & 8.7 & 6.3 & 6.2 & 4.4 &22.1 &5.3 \\
 $\mathrm{TOP}_{\text { }}$ & 12.2 & 15.4 & 12.2 & 8.5 & 5.4 & 5.1 & 4.5 &22.0 &5.2 \\
 $\mathrm{TOPG}_{ }^{\max }$ & 11.1 & 14.4 & 12.4 & 7.0 & 5.1 & 4.0 & 2.4 &19.5 & 3.5 \\
 %$\rm{POW}$ & 10.7 & 15.4 & 14.2 & 9.7 & 7.2 & 6.9 & 3.8 &18.5\\
 $2 \mathrm{G}_{\min }^{\max }$ & 10.6 & 14.3 & 11.9 & 7.7 & 5.2 & 4.2 & 1.7 &17.7& 0.8\\
 $\mathrm{TOPG}_{ }$ & 9.3 & 12.3 & 9.2 & 4.9 & 2.6 & 2.5 & 2.0 &19.3& 3.4\\
 $\mathrm{G}_{\min }^{\max }$ & 8.5 & 9.9 & 7.5 & 4.2 & 2.8 & 2.5 & 1.2 &15.2& 1.6\\
 $2 \mathrm{G}_{\min }$ & 8.5 & 12.4 & 9.4 & 5.7 & 3.1 & 2.4 & 0.6 &16.4& $-0.5$\\
 $\mathrm{SST}$ & 8.3 & 11.4 & 8.5 & 5.1 & 2.5 & 2.6 & 1.3 &17.1 & 0.1\\
 $\mathrm{G}_{\min }$ & 8.2 & 9.6 & 7.1 & 4.0 & 2.3 & 2.1 & 0.8 &14.8 & 1.0\\
 $2 \mathrm{G}^{\max }$ & 7.2 & 11.4 & 8.3 & 5.6 & 2.6 & 0.4 & $-$0.2 &16.3 & $-0.3$\\
 $2 \mathrm{G}$ & 6.8  &10.3 & 6.6 & 4.2 & 1.1 & $-$0.6 & $-$1.3 &15.1 & $-0.9$\\
 $\log \mathrm{U}$ & 5.5 & 6.0 & 5.1 & 3.5 & 3.1 & 3.2 & 1.4 &4.8& 0.5\\
 $3 \mathrm{G}$ & 5.6 & 9.5 & 5.9 & 3.8 & 0.5 & $-$1.0 & $-$1.9 &14.2& $-2.3$\\
 $\log \mathrm{N}$ & 1.8 & 3.6 & 1.2 & 1.3 & $-$0.6 & $-$1.7 & $-$1.0 &12.8& 2.2\\
 $\rm{Gamma}$ & 0.5 & 2.1 & $-0.4$ & 0.6 & $-1.1$ & $-2.4$ & $-$1.4 &11.7& $-1.8$\\

\hline \hline
\end{tabular}
 \end{center}
\end{table*}

\begin{table*}
\begin{center}
 \caption{\textbf{The priors, posteriors and 1-$\sigma$ credible intervals values for the TOP model using different data sets.}}
 \label{tab:prior_posterior}
\begin{tabular}{ccc|ccc}
\hline \hline
    \multirow{2}{*}{Model} & $ \multirow{2}{*}{\rm{Parameters} } $& $ \multirow{2}{*}{\rm{Prior} } $  &   & $ {\rm{Posterior} } $    \\
       &                  &      &  ${\mathrm{ OBS }}$ & ${\mathrm{ANA}}$ &$ {\mathrm{PHE}}$  \\ 
\hline 
           $ $    & $ m_{\mathrm{min }} [\mathrm{M_{\odot}}] $ & $ U[0.9,1.5] $ & $ 1.08^{+0.04}_{-0.04} $ & $ 1.10^{+0.04}_{-0.05} $ & $1.10^{+0.04}_{-0.06}$ \\
$\mathrm{TOP}$ & $\delta_{m} [\mathrm{M_{\odot}}]$& $U[0.01,1]   $               & $0.32^{+0.15}_{-0.12} $& $0.24^{+0.14}_{-0.11} $& $0.17^{+0.13}_{-0.10}$\\
               & $m_{\mathrm{max }} [\mathrm{M_{\odot}}] $& $U[1.5,2.9]$ & $2.38^{+0.23}_{-0.13}$ & $2.36^{+0.29}_{-0.17}$ & $2.23^{+0.33}_{-0.15}$ \\
               & $\alpha$ & $U[-5,25] $                                          & $5.64^{+1.18}_{-1.09}$ & $6.47^{+1.28}_{-1.15} $& $6.28^{+1.23}_{-1.19}$\\
\hline \hline
\end{tabular}
 \end{center}
\end{table*}

\clearpage

\begin{figure}
\centering
\includegraphics[width=120mm]{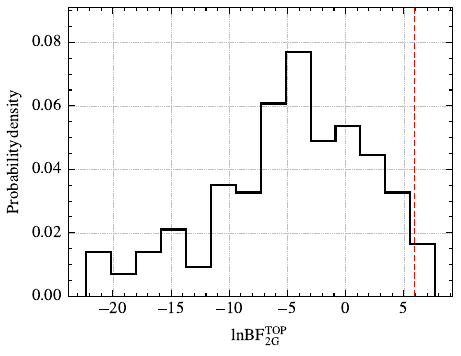}\\
\caption{The distribution of log Bayes factor (between the TOP and 2G model) for a simulated data set that includes 87 neutron star mass measurements drawn from a two-Gaussian distribution. The red dashed line is ${\rm ln BF^{TOP}_{2G}}=5.8$ computed for the actual ANA data set. This plot shows that the chance to falsely reject the two-Gaussian model for current observations of neutron stars is small, about $2.5\%$.}
\label{fig:BF_hist}
\end{figure}

\clearpage

\begin{figure}
\centering
\includegraphics[width=120mm]{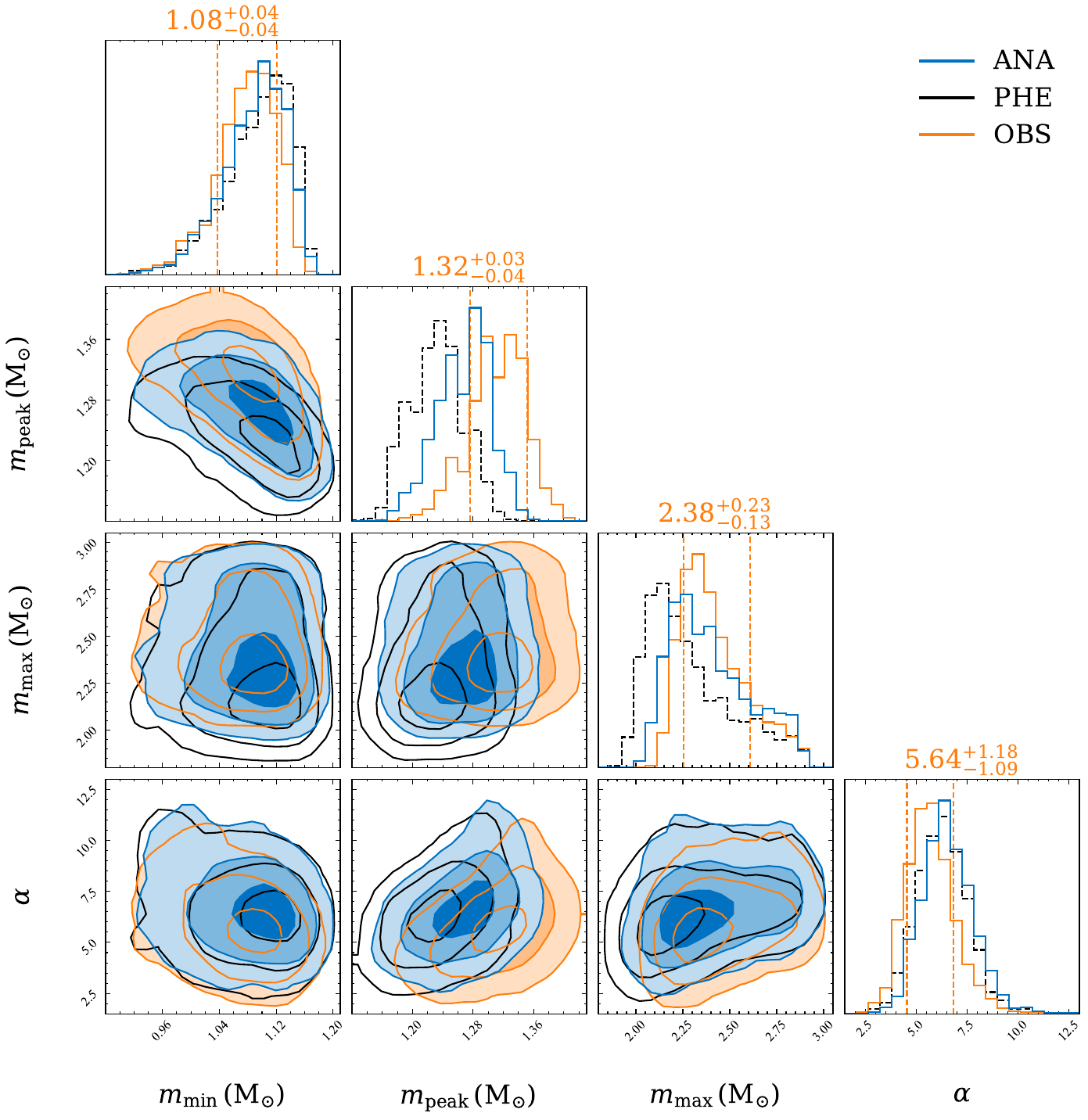}\\
\caption{Posterior distribution of the TOP model parameters for the OBS (orange), ANA (blue) and PHE (black) data set. The vertical lines mark the 1-$\sigma$ credible interval of the marginalized posteriors for the OBS data set.}
\label{fig:TOP_corner}
\end{figure}

\clearpage

\begin{figure}
\centering
\includegraphics[width=92mm]{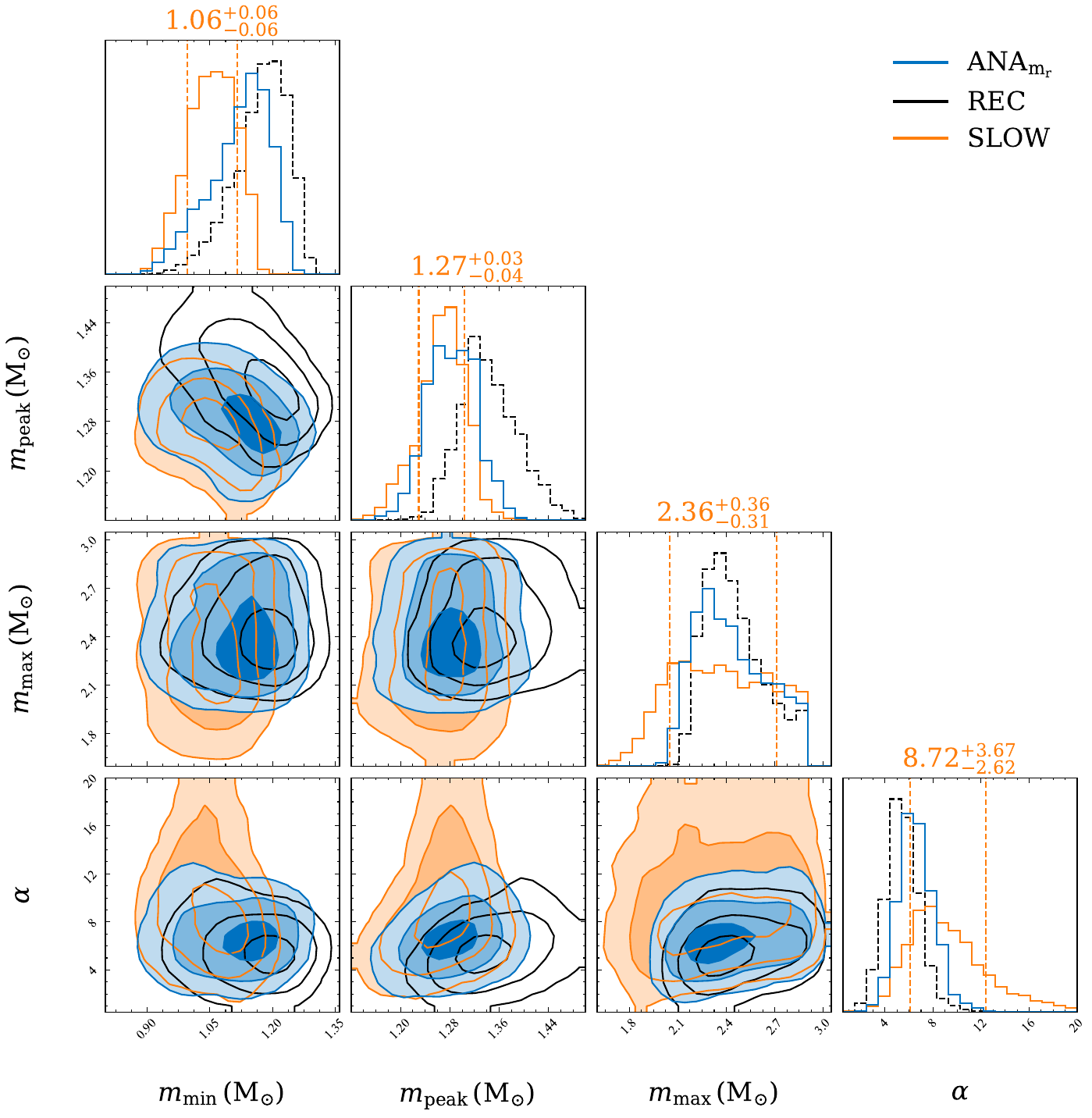}
\includegraphics[width=92mm]{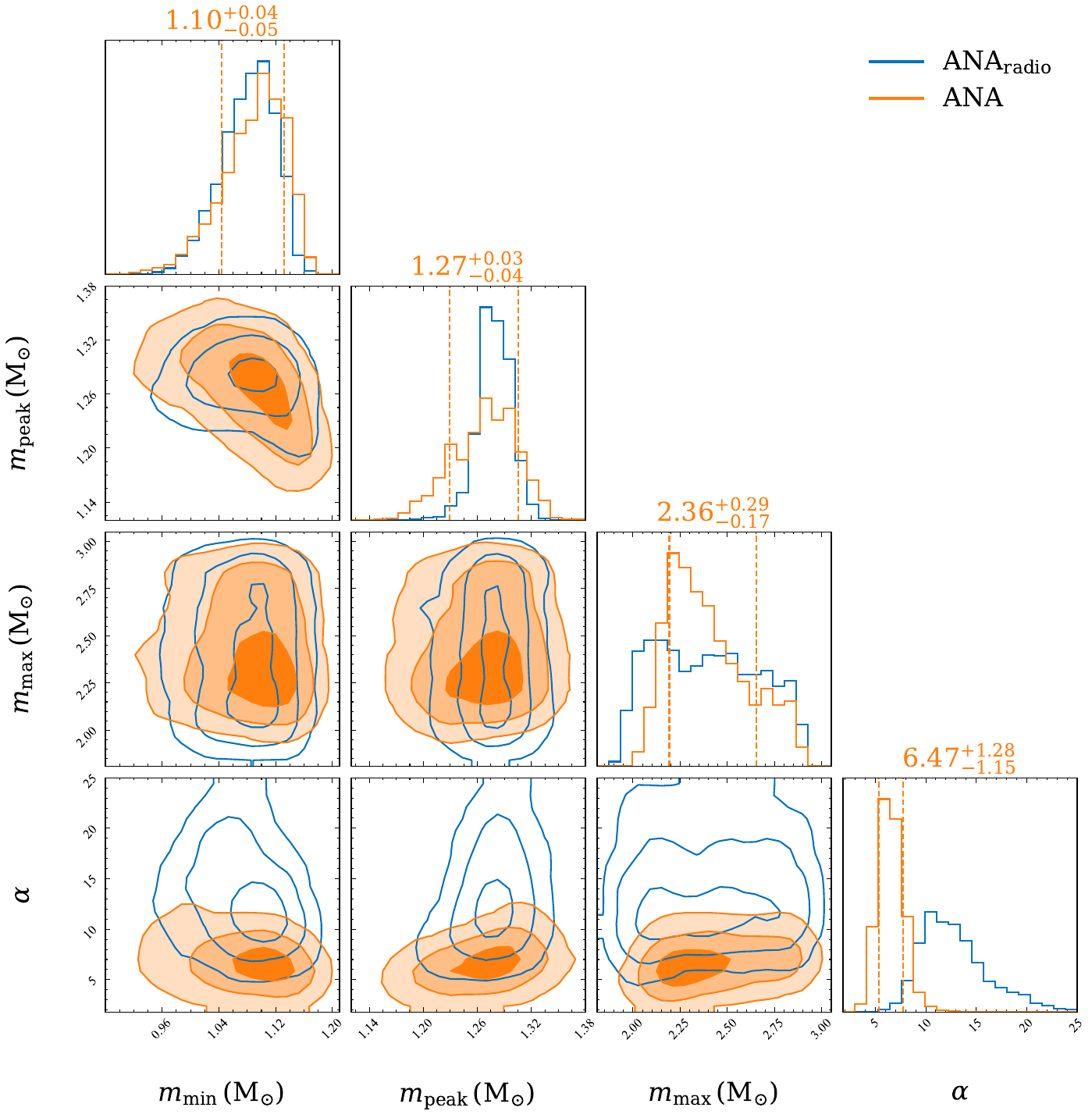}
\\
\caption{As Extended Data Figure \ref{fig:TOP_corner}, but for the ANA$_{m_{r}}$ (blue), SLOW (orange) and REC (black) data set on the top panel; and on the bottom panel for the ANA (orange) and ANA$_{\rm {radio}}$ (blue) which is a subset of ANA by only including mass measurements from radio pulsar observations.}
\label{fig:TOP_corner1}
\end{figure}

\section{Semi-analytic supernova modeling}
\label{sec:supernova_modeling}

In this section we investigate the implications of the measured neutron star birth mass distribution for core-collapse supernova physics.
%This is illustrated by Figure~\ref{fig:ns_mass_predict}. 
Extended Data Figure~\ref{fig:ns_mass_predict}a shows the distribution $\phi(M_4)$ of the iron-silicon core mass $M_4$ (defined by a
threshold value of $4\,\mathrm{k}_\mathrm{B}$ per nucleon in specific entropy), weighted by the initial mass function (IMF) for massive stars,\cite{Kroupa01IMF,Schneider18IMF} for a set of 2912 single-star supernova progenitor models\cite{MHLC_16} with initial masses from $\unit[9.45]{\mathrm{M}_{\odot}}$ to $\unit[45]{\mathrm{M}_{\odot}}$.
% $9.45 M_\odot$ to $45 M_\odot$. %(which extends the progenitor set of ).
The parameter $M_4$ is a proxy for the mass coordinate for shock revival in neutrino-driven supernovae \cite{ertl_16}. 
%EHT: An additional factor $(M_4/M_0)^{6.5}$ (with an arbitrary scaling factor $M_0$) is applied to highlight deviations from a power law with slope $-6.5$. 
An additional factor $(M_4/M_0)^{6.5}$ (with an arbitrary scaling factor $M_0$) is applied to highlight deviations from the fiducial $\phi(M_4) \propto M_4^{-6.5}$ scaling where the power-law slope comes from the TOP model as determined for the birth mass function (see Figure \ref{fig:TOP_m_peak} in the main text). 

Without additional accretion after shock revival, the neutron star mass distribution would be expected to resemble the distribution of $M_4$ because the neutron star mass is expected to scale linearly with $M_4$.
Extended Data Figure~\ref{fig:ns_mass_predict}a shows that this distribution not only shows multiple peaks, but is also top-heavy compared to  the observed distribution of neutron star mass. 
The peaked nature of the theoretical neutron star mass distribution is found to persist in binary population synthesis simulations utilising modern semi-analytic supernova prescriptions\cite{Vigna_18}.

In order to flatten $\phi (M_4) (M_4/M_0)^{6.5}$ (to make it look more like the distribution of observed neutron star masses), one either needs to remove neutron stars from the massive end of the distribution, or distribute neutron stars in the peaks more broadly towards higher masses. 
Applying the semi-analytic supernova model of Ref. \cite{MHLC_16} with default parameters, the resulting neutron star mass distribution deviates less from the $M^{-6.5}$ power law because the model predicts most stars above $\unit[20]{\mathrm{M}_{\odot}}$ to collapse to black holes (Extended Data Figure~\ref{fig:ns_mass_predict}b).
% Extended Data Figure~\ref{fig:ns_mass_predict2} shows $\phi(M)$ without the compensation factor $(M/M_0)^{6.5}$). 
The peaks from the distribution of $M_4$ are still present, however, and a new peak around $\unit[1.8]{\mathrm{M}_{\odot}}$ emerges because shock revival does not occur at the edge of iron-silicon core, but at a mass shell further out. 

In order to improve the agreement between the semi-analytic supernova model and the observed distribution, we adjust the semi-analytic model to increase accretion after shock revival by reducing the shock compression ratio to $\beta=3.1$, and to reduce the mass range for successful explosions while maintaining realistic explosion properties, as shown in Extended Data Figure~\ref{fig:expl_modified}.
We choose a turbulent shock expansion multiplier 
$\alpha_\mathrm{turb}=1.10$, an efficiency $\zeta=0.85$ for the neutrino accretion luminosity and a cooling time scale $\tau_{1.5}=1.2\,\mathrm{s}$ for a $\unit[1.5]{\mathrm{M}_{\odot}}$ neutron star.  
This broadens the peak at $1.5 M_\odot$ in the standard model and shifts it to higher masses, bringing the mass distribution close to an $M^{-6.5}$ power law up to $M = \unit[1.9]{\mathrm{M}_{\odot}}$ 
(Extended Data Figure~\ref{fig:ns_mass_predict}c and Extended Data Figure~\ref{fig:ns_mass_predict2}c). 

The trough between low-mass neutron stars from stars with $< \unit[12]{\mathrm{M}_{\odot}}$
is largely filled by a cluster of exploding progenitors around $\unit[17]{\mathrm{M}_{\odot}}$ (Extended Data Figure~\ref{fig:expl_modified}c).
This is accomplished by allowing neutron stars from
$12\texttt{-}15 M_\odot$ to accrete varying amounts of mass up to $\unit[0.14]{\mathrm{M}_{\odot}}$ after shock revival (Extended Data Figure~\ref{fig:ns_mass_predict2}). 
The predicted mass distribution still has (less prominent) troughs and a pile-up of neutron stars at $2 M_\odot$, but these can be ascribed to imperfections in the semi-analytic supernova model. 
For light neutron stars, the model predicts unrealistically high net mass loss of up to $\unit[0.04]{\mathrm{M}_{\odot}}$ from the neutron star after early shock revival, which is not seen in multi-dimensional simulations. 
Under realistic conditions neutron stars from progenitors with $< \unit[12]{\mathrm{M}_{\odot}}$ should therefore have higher masses and could fill the trough at $1.25\texttt{-}1.4 M_\odot$ in the distribution.
The excess of very massive neutron stars around $\unit[2]{\mathrm{M}_{\odot}}$ could be removed in nature if roughly half of these neutron stars undergo more fallback to form black holes instead, or if many of the corresponding progenitors do not actually explode.

% \bm{TO DO: Based on Alex' and Simon's arguments, we need to explain why we ruled out other explanations}

\clearpage

\begin{figure}%[htp]
\centering
% \vspace{-20mm}
\textbf{a}\includegraphics[width=.40\textwidth]{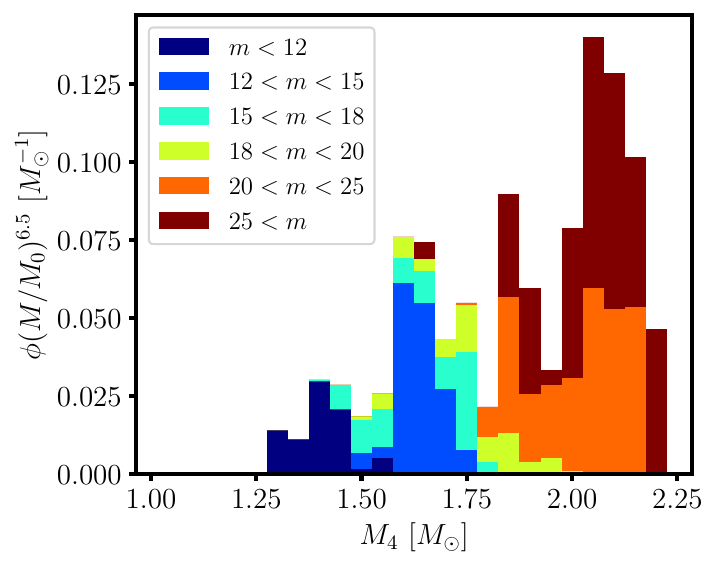}\\ \vspace{2mm}
\textbf{b}\includegraphics[width=.40\textwidth]{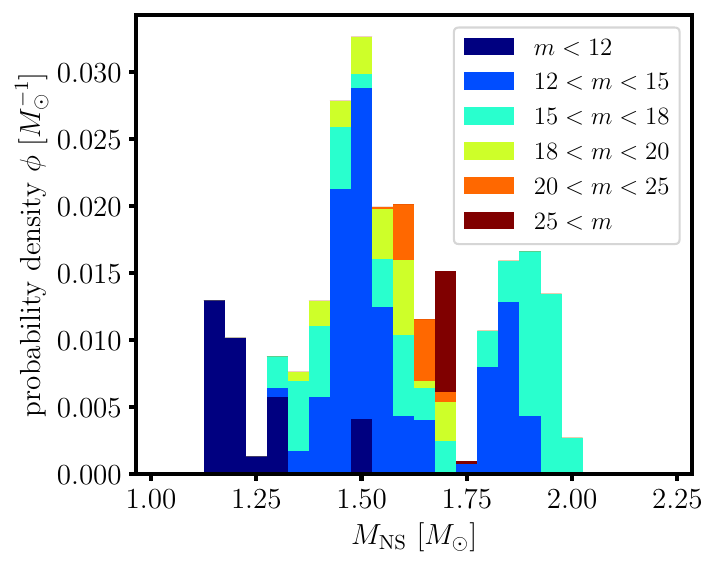}\\ \vspace{2mm}
\textbf{c}\includegraphics[width=.40\textwidth]{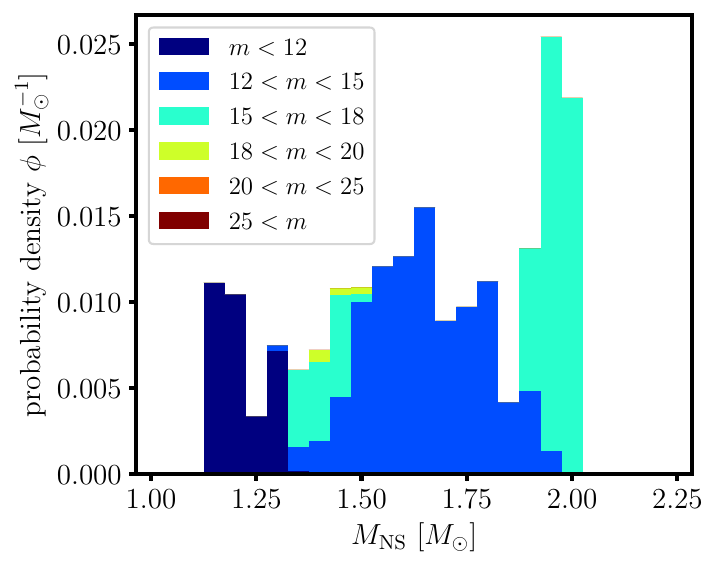} \\
\vspace{5mm}
\caption{\textbf{a}, IMF-weighted distribution $\phi(M_4)$ of the mass $M_4$ of the iron-silicon core for a set of 2912 single-star supernova progenitors of solar metallicity\cite{MHLC_16}, with a compensation factor $(M_4/M_0)^{6.5}$ to show deviations from the power-law of the measured neutron star mass distribution. 
\textbf{b}, Predicted distribution $\phi(M)$ of neutron star masses with the same compensation factor based on the semi-analytic supernova model of ref.\cite{MHLC_16} with standard model parameters.
\textbf{c}, Predicted distribution $\phi(M)$ of neutron star masses with modified supernova model parameters to allow for more accretion after shock revival.
}
\label{fig:ns_mass_predict}
\end{figure}

\clearpage

\begin{figure}%[htp]
\centering
\textbf{a}\includegraphics[width=.42\textwidth]{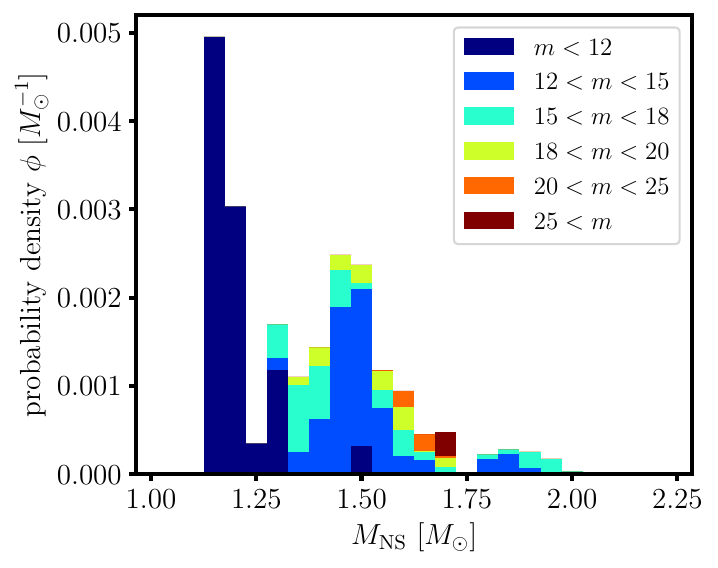}
\textbf{b}\includegraphics[width=.42\textwidth]{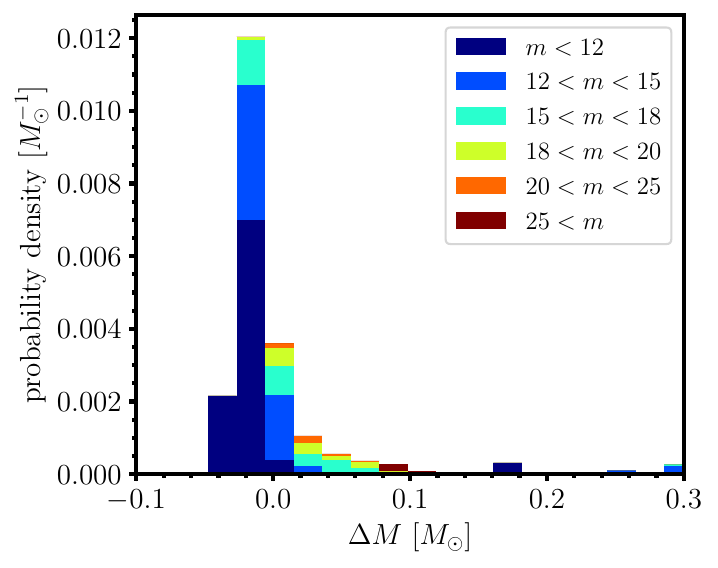}
\\
\textbf{c}\includegraphics[width=.42\textwidth]{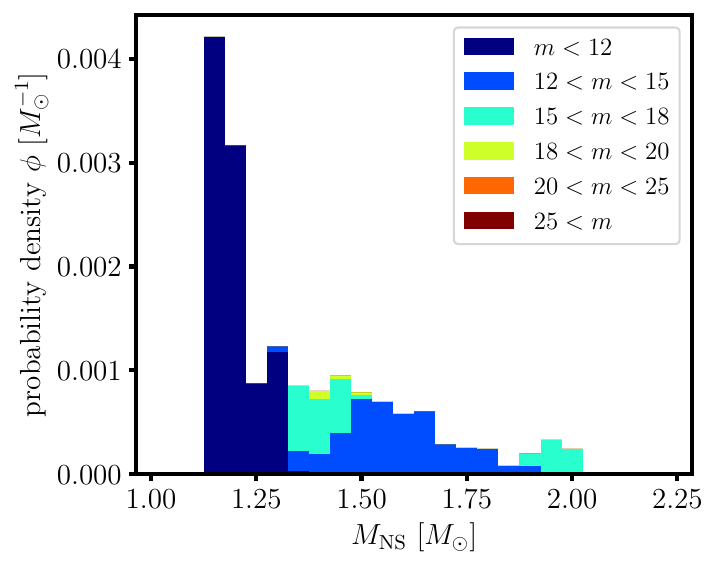}
\textbf{d}\includegraphics[width=.42\textwidth]{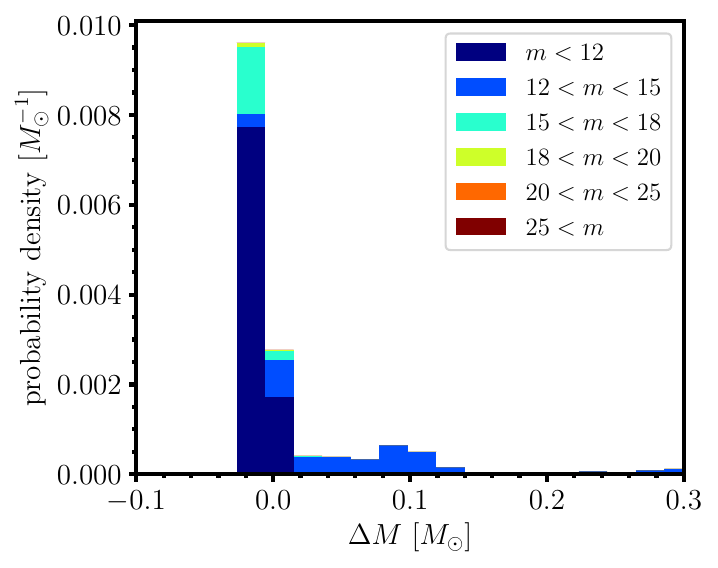} \\
\vspace{10mm}
\caption{
\textbf{a}, Predicted IMF-weighted neutron star mass distribution
$\phi(M)$ using the semi-analytic supernova model of ref.\cite{MHLC_16} with standard model parameters. \textbf{b}, Distribution of the amount $\Delta M$ of net accretion after shock revival.
Panels \textbf{c} and \textbf{d} Predicted IMF-weighted distribution
of neutron star masses and accretion after shock revival for the modified supernova model with enhanced accretion during the explosion phase.
}
\label{fig:ns_mass_predict2}
\end{figure}

\clearpage

\begin{figure}%[htp]
\centering
\includegraphics[width=\textwidth]{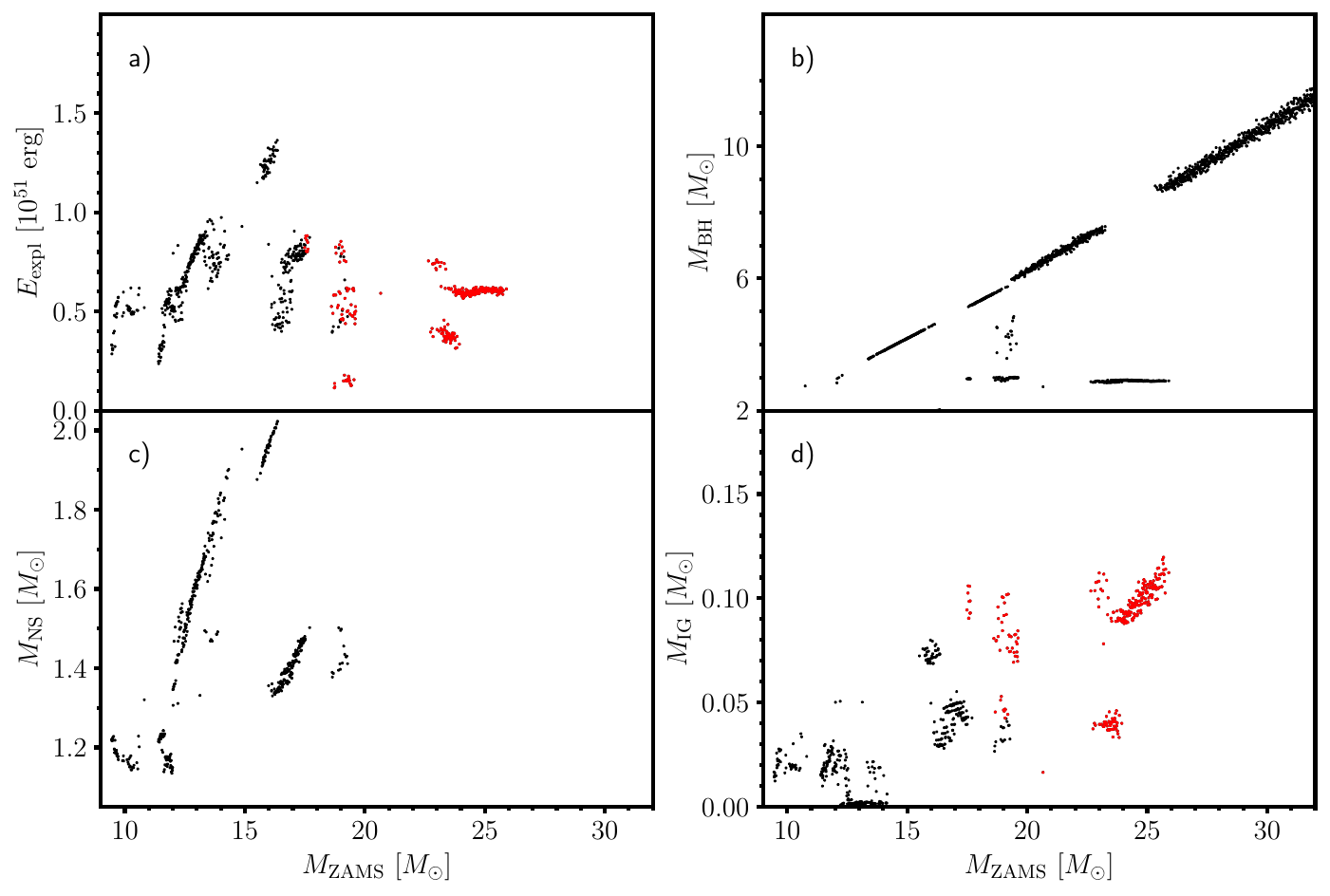} \\
\vspace{5mm}
\caption{\textbf{a}) Explosion energy, \textbf{b}) black-hole masses for non-exploding cases and fallback supernovae, \textbf{c}) neutron star masses, \textbf{d}) nickel masses predicted by the
semi-analytic supernova model with adjusted parameters
for enhanced accretion to better match the observed power-law distribution of neutron star masses. Red dots in panels \textbf{a} and \textbf{d} denote explosions with black hole formation by fallback. Explosion energies and nickel masses span the typical range observed for Type~IIP supernovae.
}
\label{fig:expl_modified}
\end{figure}

\section{Potential selection effects}
\label{sec:disc}

Here we discuss potential selection effects that might have had an impact on the inference of neutron star mass distribution.
Our compilation of neutron-star mass measurements was the result of a diverse range of observations, which may have introduced different observation biases.
For simplicity, we first focus on our main subset of data: masses measured with the radio pulsar timing method.
Extended Data Figure \ref{fig:mass-spin}\textbf{a} shows the measured mass versus the observed spin period for 39 recycled pulsars (blue) and other neutron stars (orange).
Extended Data Figure \ref{fig:mass-spin}\textbf{b} shows the measured mass versus the luminosity at 1.4 GHz calculated from the flux densities and pulsar distances as reported in the ATNF Pulsar Catalogue\cite{PSRcat05} for 39 recycled pulsars.
From both plots, one can see that there is no apparent correlation between the mass and the detectability of a pulsar.
This indicates that our inferred neutron-star mass distribution is unlikely to be biased by selection effects associated with pulsar surveys.

Another potential selection bias may have arisen because all but one of neutron stars in our compilation are in binary systems\footnote{The one isolated neutron star (PSR J0030+0451) is a fully recycled pulsar. It was probably part of a disrupted binary system.}.
As massive stars, the progenitors of neutron stars, are commonly formed in binary systems, the neutron star mass function determined in this work may be a good representation of the overall neutron star population, except that binary systems containing neutron stars seen today are only a fraction of binary systems that have given birth to neutron stars.
The supernova explosion associated with the formation of a neutron star could disrupt the binary system.
If the survival probability of binary systems is correlated with the neutron star mass, for example, more massive neutron stars receive larger kicks at birth, then the population of observed binary pulsar systems would be biased towards a particular range of neutron star masses.
The effect of such a correlation between the neutron star mass and the supernova kick has been tentatively investigated in population synthesis calculations\cite{Vigna_18,Schneider_21}. 
In particular, using neutron star birth masses derived from modern, parameterized supernova models, ref.\cite{Vigna_18} found a neutron-star birth-mass distribution for the primary in double neutron star binaries that is skewed towards lower masses than the single-star mass function. 
Conversely, the mass distribution of secondaries is shifted upwards slightly compared to the single-star case
(see their figure~7). 
Although the aggregated distribution of neutron star birth masses, in particular its shape, is like that of the single-star case, systematic differences in the slope of the mass distribution in double neutron stars, neutron star-white dwarf binaries and isolated neutron stars cannot be excluded; see, for example, figure~8a in ref.\cite{Schneider_21}.
This will be an interesting area for future research with a much larger catalogue of neutron star mass measurements.

To summarize, although we believe that our results are robust against potential selection effects, we do expect to unveil a fuller picture of neutron-star mass function in the next decade or so when we have a sizeable sample of mass measurements for every subpopulation of neutron star systems, for example, binary neutron star mergers from gravitational-wave observations, neutron star and white dwarf binaries from pulsar observations, and isolated neutron stars from microlensing observations\cite{DaiShi15,Mandel20lens}.

\clearpage

\begin{figure}
\centering
\includegraphics[width=94mm]{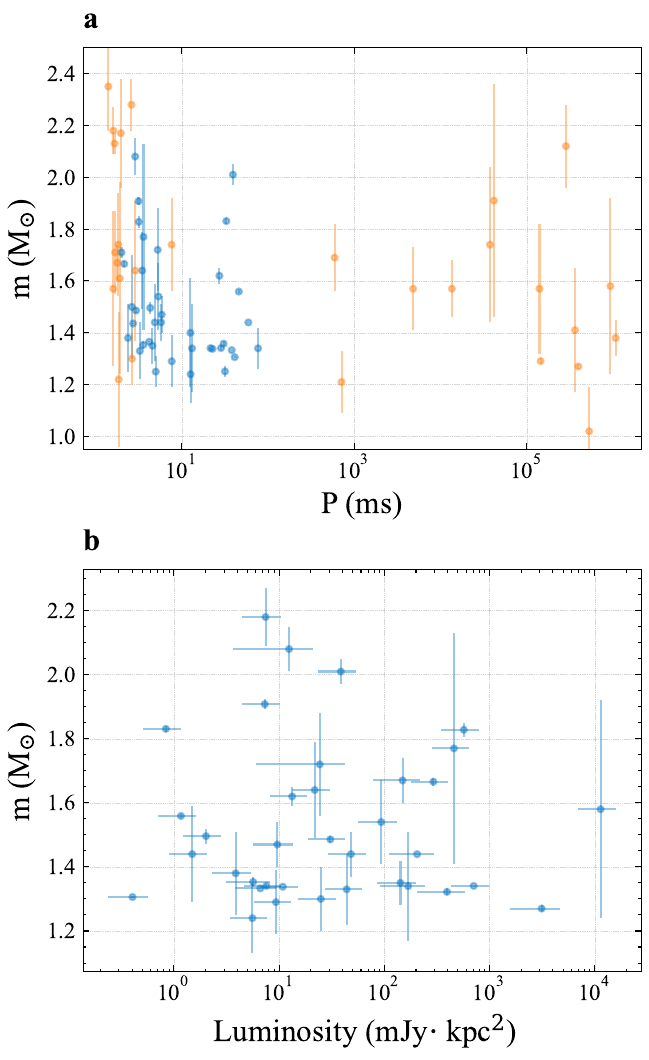}
\\
\caption{\textbf{The period and luminosity distribution of pulsars with measured mass.} \textbf{a}, Measured mass versus spin period for 39 recycled 
pulsars (blue) in group $a$ ("Modelling the accreted masses of recycled pulsars` 
in Methods) and the remaining neutron stars (orange).
\textbf{b}, Measured mass versus luminosity at 1.4 GHz for 39 recycled pulsars, where the horizontal error bars account for uncertainties in the flux density and distance. We adopt the distance estimates from the Australia Telescope National Facility Pulsar Catalogue\footnote{\url{https://www.atnf.csiro.au/research/pulsar/psrcat/}} while assuming a 20\% error. In both panels, plotted are mean values with 1$\sigma$ credible errors, while the measurement uncertainty of spin periods in the upper panel is too small to be seen}
\label{fig:mass-spin}
\end{figure}

\end{methods}

% \bibliographystyle{plain}

%\bibliography{AlignedIsotropicGW}
%\bibliographystyle{Science}
% \bibliography{ref.bib}

\end{document}